\documentclass{egpubl4arxiv}

\JournalSubmission    
\usepackage[T1]{fontenc}
\usepackage{dfadobe}  

\biberVersion
\BibtexOrBiblatex
\usepackage[backend=biber,bibstyle=EG,citestyle=alphabetic,backref=true,url=false,bibencoding=utf8]{biblatex} 
\electronicVersion
\PrintedOrElectronic

\ifpdf \usepackage[pdftex]{graphicx} \pdfcompresslevel=9
\else \usepackage[dvips]{graphicx} \fi

\usepackage{egweblnk}
\usepackage[final,tracking=true,kerning=true,spacing=true,factor=1100,stretch=10,shrink=10]{microtype}
\SetTracking{encoding=*, shape=sc}{15}

\usepackage{booktabs} 
\usepackage{newtxmath}  
\usepackage{rotating}

\usepackage[dvipsnames]{xcolor}

\newcommand{\mbf}{\mathbf}

\usepackage{float}
\usepackage{amsmath}
\usepackage{gensymb}
\usepackage{textcomp}
\usepackage{etoolbox}
\usepackage{xspace}

\makeatletter
\DeclareRobustCommand\onedot{\futurelet\@let@token\@onedot}
\def\@onedot{\ifx\@let@token.\else.\null\fi\xspace}

\def\ie{\emph{i.e}\onedot}

\def\etal{\emph{et al}\onedot}
\makeatother

\usepackage{enumitem} 
\usepackage{subcaption}
\usepackage{caption}

\title
      {Neural BRDF Representation and Importance Sampling}

\author[A. Sztrajman, G. Rainer, T. Ritschel, T. Weyrich]
{
	\parbox{\textwidth}{\centering A. Sztrajman
	\quad G. Rainer \quad T. Ritschel \quad T. Weyrich} \\
	University College London, London, UK
}

\begin{document}


\maketitle
\begin{abstract}
Controlled capture of real-world material appearance yields tabulated sets of highly realistic reflectance data. In practice, however, its high memory footprint requires compressing into a representation that can be used efficiently in rendering while remaining faithful to the original. Previous works in appearance encoding often prioritised one of these requirements at the expense of the other, by either applying high-fidelity array compression strategies not suited for efficient queries during rendering, or by fitting a compact analytic model that lacks expressiveness. We present a compact neural network-based representation of BRDF data that combines high-accuracy reconstruction with efficient practical rendering via built-in interpolation of reflectance. We encode BRDFs as lightweight networks, and propose a training scheme with adaptive angular sampling, critical for the accurate reconstruction of specular highlights. Additionally, we propose a novel approach to make our representation amenable to importance sampling: rather than inverting the trained networks, we learn 
to encode them in a more compact embedding that can be mapped to parameters of an analytic BRDF for which importance sampling is known. We evaluate encoding results on isotropic and anisotropic BRDFs from multiple real-world datasets, and importance sampling performance for isotropic BRDFs mapped to two different analytic models.
\end{abstract}  

\section{Introduction}
\label{sec:intro}

Accurate reproduction of material appearance is a major challenge in
computer graphics. Currently, there are no standardised representations
for reflectance acquisition data, and there is no universal analytic
model capable of representing the full range of real-world
materials~\cite{brdfstar2016}.
The development of new methods for appearance capture has led to an
increasing amount of densely sampled data from real-world
appearance~\cite{mpbm03, utiadensebrdf, rgl}. Although tabulated
representations of reflectance data are usually very accurate, they
suffer from a high memory footprint and computational cost at
evaluation time~\cite{deepbrdf}. Reflectance data, however, exhibits strong
coherence~\cite{dong2019}, which can be leveraged for efficient
representation and evaluation of real-world materials. Existing
approaches perform dimensionality reduction using
matrix factorisation~\cite{lawrence2004, ngan2006, nielsen2015}
which requires a large number of
components for high quality reproduction, or by fitting analytic
models~\cite{ngan2005}, usually relying on time-consuming and
numerically unstable nonlinear optimisation and presenting a
limited capacity to accurately reproduce real-world materials.

Recent works successfully applied deep learning methods on
reflectance estimation~\cite{deschaintre2018}, material
synthesis~\cite{zsolnai2018} and BTF compression and
interpolation~\cite{gilles2019, gilles2020}. Close to our work,
Hu~\etal's DeepBRDF~\cite{deepbrdf} use a deep convolutional
autoencoder to generate compressed encodings of measured BRDFs, which
can be used for material estimation and editing; however, their
encoding depends on a rigid sampling of the tabulated data,
independent of the shape of the encoded BRDF,
and DeepBRDFs require back-transformation
into tabulated form for evaluation, making
them less suitable for rendering than for editing of appearance.

In contrast, we aim for a representation that allows for efficient
rendering while retaining sufficient expressiveness for a wide range
of materials.
The contributions of our work are as follows:
\begin{itemize}
    \item A neural representation for measured BRDFs that
    \begin{itemize}
        \item retains high fidelity under a high compression rate;
        \item can be trained with an arbitrary sampling of the original BRDF, allowing for BRDF-aware adaptive sampling of the specular highlights during training which is critical for their accurate reconstruction; additionally, our network
        \item can be used directly as replacement of a BRDF in a rendering pipeline, providing built-in evaluation and interpolation of reflectance values, with speeds comparable to fast analytic models. In Sections~\ref{sec:res:reconstruction}, \ref{sec:res:anisotropic} and~\ref{sec:res:performance} we compare our encoding with other representations in terms of quality of reconstruction, speed and memory usage.
      \end{itemize}
    \item Deployment of a learning-to-learn autoencoder architecture
    to explore the subspace of real-world materials by learning a
    latent representation of our Neural-BRDFs (NBRDFs). This
    enables further compression of BRDF data to a $32$-values
    encoding, which can be smoothly interpolated to create new
    realistic materials, as shown in
    Section~\ref{sec:res:newmaterials}.
    \item A learned mapping between our neural representation and an invertible
    parametric approximation of the BRDF, enabling importance
    sampling of NBRDFs in a rendering pipeline; in
    Section~\ref{sec:res:sampling} we compare our method with other
    sampling strategies.
\end{itemize}

\section{Related Work}
\label{sec:rel}

\subsection{BRDF Compression and Interpolation}
\label{sec:rel:compression}

Real-world captured material appearance is commonly represented by
densely sampled and high-dimensional tabulated BRDF
measurements. Usage and editing of these representations usually
requires strategies for dimensionality reduction,
most commonly
through different variants of matrix factorisation~\cite{lawrence2004, ngan2006, nielsen2015},
which require large storage in order to provide accurate reconstructions,
or by fitting to an analytic model. BRDF models are lightweight
approximations specifically designed for compact representation and efficient
evaluation of reflectance data. However, fitting these models usually
relies on unstable optimisations, and they are capable of
representing a limited gamut of real-world appearances~\cite{sztrajman2017}.

Ngan~\etal~\cite{ngan2005} were the first to systematically study the fitting
of analytical BRDF models to real-world materials. Since then, more complex
models have been developed, many of them based on the microfacet model originally
proposed by Cook and Torrance~\cite{cooktorrance1982}. In particular, two parameterisations
of the microfacet $D$ distribution are considered the state-of-the-art in parametric reconstruction: the shifted gamma distribution (SGD) by Bagher~\etal~\cite{bagher2012} and the ABC model by Low~\etal~\cite{low2012}.

More recent models have been developed with non-parametric definitions of some or all
component functions of the microfacet model. 
Although these models are limited by their inherent factorisation assumptions,
they present a very good trade-off between memory storage and high-quality reconstruction.
Dupuy~\etal~\cite{dupuy2015} fit the $D$
distribution from the retro-reflective lobe using power iterations. Their fitting method
avoids the instabilities of nonlinear optimisation and allows the subsequent translation
to other microfacet-based models such as GGX~\cite{walter2007microfacet} and Cook-Torrance~\cite{cooktorrance1982}. Bagher~\etal~\cite{bagher2016} define a non-parametric factor microfacet model (NPF), state-of-the-art in non-parametric reconstruction of isotropic BRDF, using tabulated definitions for the three functional components ($D$, $F$ and $G$) of the microfacet model, with a total
memory footprint of $3.2$KB per material. Dupuy and Jakob~\cite{rgl} define a new adaptive
parameterisation that warps the 4D angle domain to match the shape of the material. This
allows them to create a compact data-driven representation of isotropic and anisotropic
reflectance. Their reconstructions compare favorably against NPF, although at the price
of an increased storage requirement ($48$KB for isotropic 3-channels materials, $1632$KB for anisotropic).

Close to our work, Hu~\etal~\cite{deepbrdf} use a convolutional
autoencoder to generate compressed embeddings of real-world BRDFs,
showcasing applications on material capture and editing.
In Section~\ref{sec:met:nbrdf}
we describe a method for BRDF compression
based on a neural representation of material appearance. In
contrast with Hu~\etal's, our neural BRDF network can be directly used
as replacement of a BRDF in a rendering system, without the need to
expand its encoding into a tabular representation. Moreover, NBRDF provides
built-in fast interpolated evaluation, matching the speed of analytic
models of much lower reconstruction quality. We compare our method with 
other parametric and non-parametric representations in terms of reconstruction
accuracy, compression and evaluation speed.

Chen~\etal~\cite{chen2020} implement iBRDF, a normalising flow network 
designed to encode reflectance data, focusing on generating a differentiable
inverse rendering pipeline for joint material and lighting estimation 
from a single picture with known geometry.
Their architecture, based on non-linear independent components
estimation (NICE)~\cite{dinh2015}, compares favourably against
bi-variate tabulated representations~\cite{romeiro2008} of the
MERL BRDF database~\cite{mpbm03} (detailed in Section~\ref{sec:met:merl}) in terms of reconstruction accuracy, with similar storage requirements.
Similarly to our architecture, the input of iBRDF is given by the Rusinkiewicz
parameterisation~\cite{rusinkiewicz1998}, although it is reduced to three angles,
thus limiting the representation to isotropic materials.

In Section~\ref{sec:met:ae}, we describe a learning-to-learn
autoencoder architecture that is able to further compress our NBRDF
networks into a low dimensional embedding. A similar architecture was
previously used by Maximov~\etal~\cite{dam} to encode deep appearance
maps, a representation of material appearance with baked scene
illumination. Soler~\etal~\cite{soler2018} explored a low-dimensional
nonlinear BRDF representation via a Gaussian process model, supporting
smooth transitions across BRDFs. Similarly,
in Section~\ref{sec:res:newmaterials} we show that the low
dimensional embeddings generated by our autoencoder can be
interpolated to create new realistic materials.

\subsection{Importance Sampling of Reflectance Functions}

BRDF-based importance sampling is a common strategy used to reduce the
variance of rendering algorithms relying on Monte Carlo
integration~\cite{siggraph2010course}. For some analytic BRDF models,
such as Blinn-Phong~\cite{blinn77}, Ward~\cite{ward1992},
Lafortune~\cite{lafortune1997} and
Ashikhmin-Shirley~\cite{ashikhmin2000anisotropic}, it is possible to
compute the inverse cumulative distribution function analytically,
thus providing a fast method for importance sampling. For the general
case, however, closed-form inverse CDFs do not exist, requiring costly
numerical calculation.

A practical alternative is to approximate the original BRDF by a PDF
with a closed-form inverse CDF, and to use them for importance
sampling instead~\cite{lawrence2004}. While generally sacrificing
speed of convergence, this approach still leads to accurate, unbiased
results in the limit; however, it often introduces the requirement of
a potentially unreliable
model fit via non-linear optimisation.
Accordingly, in the context of measured data, many works forgo
non-linear models in favour of numerically more robust approximations,
including matrix factorisation~\cite{lawrence2004}, as well as
wavelets~\cite{clarberg2005} and spherical harmonics
approximations~\cite{jarosz2009}.
Our work, too, operates with an approximating PDF, but retains a
physically-based, non-linear invertable model and 
eliminates the non-linear optimisation
by training a fast neural network to fit the model
parameters to measured BRDF data (see Section~\ref{sec:met:sampling}).

\subsection{Neural Sampling and Denoising}

While importance sampling's main objective is faster convergence, it
has the secondary effect of reducing noise. Convolutional networks
have successfully been applied for denoising of Monte Carlo
renderings~\cite{chakravarty2017, bako2017} and radiance interpolation
from sparse samples~\cite{ren2013, kallweit2017}. However these
methods do not converge to ground truth, since they act directly on
rendered images, lacking information from the underlying scene.

Other recent works, too, leveraged neural networks for importance sampling in Monte Carlo integration. Bako~\etal~\cite{bako2019} trained a deep convolutional neural network for guiding path tracing with as little as 1 sample per pixel. Instead of costly online learning, their offline deep importance sampler (ODIS) is trained previously with a dataset of scenes and can be incorporated into rendering pipelines without the need for re-training. 
Lindell~\etal~\cite{lindell2021} developed AutoInt, a procedure for fast evaluation of integrals based on a neural architecture trained to match function gradients. The network is reassembled to obtain the antiderivative, which they use to accelerate the computation of volume rendering.
Zheng~\etal~\cite{zheng2019} trained an invertible real-valued non-volume preserving network (RealNVP)~\cite{dinh2017} to generate a scene-dependent importance sampler in primary sample space. Concurrently, M\"uller~\etal~\cite{mueller2019} trained an invertible neural network architecture based on non-linear independent components estimation (NICE)~\cite{dinh2015} for efficient generation of samples. They explored two different settings: a global offline high-dimensional sampling in primary sample space, and a local online sampling in world-space, applied to both incident-radiance-based and product-based importance sampling. An additional network is used to learn approximately optimal selection probability and further reduce variance.

\section{Method and Implementation}
\label{sec:met}

Drawing upon the observations of Section~\ref{sec:rel}, we propose a
new representation for measured BRDFs that maximises fidelity to the
data while retaining practicality. The remainder describes our basic
reflectance encoding (Section~\ref{sec:met:nbrdf}), an auto-encoder framework for efficient representation (Section~\ref{sec:met:ae}), 
as well as
an importance sampling scheme to further
speed-up rendering (Section~\ref{sec:met:sampling}).

\subsection{BRDF Encoding}
\label{sec:met:nbrdf}

\begin{figure}[t]
    \centering
    \includegraphics[width=0.7\linewidth]{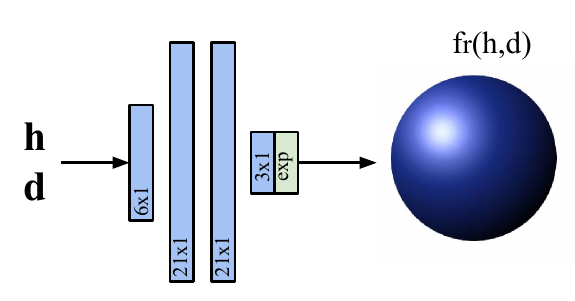}
    \caption{\label{fig:nbrdf} Diagram of a Neural-BRDF (NBRDF).}
\end{figure}

Our representation for BRDF data uses a shallow fully-connected
network with ReLU activations and a final exponential layer,
as shown in Figure~\ref{fig:nbrdf},
which we will refer to as NBRDF (Neural-BRDF).
These NBRDFs work as a standard BRDF representation for a single material: 
the network takes incoming and outgoing light directions as input,
and outputs the associated RGB reflectance value.
Interpolation is handled implicitly by the network, via the continuous input space.

\begin{figure}[b]
    \centering
    \includegraphics[width=\linewidth]{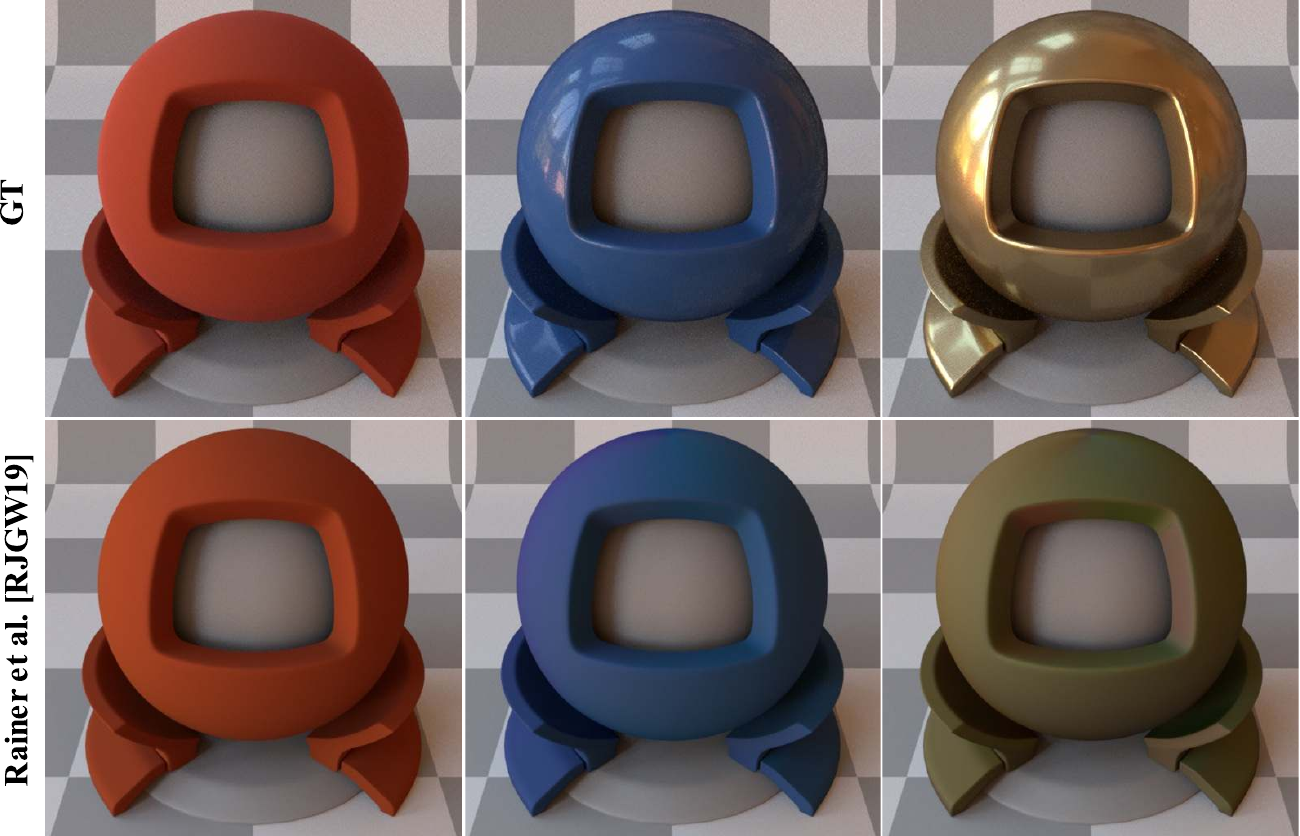}
    \caption{\label{fig:btfnet} Top row: Ground truth. Bottom row: Reconstruction using Rainer~\etal's architecture~\cite{gilles2019}, treating each BRDF as a spatially uniform BTF.}
\end{figure}

The parametrisation of the network input strongly affects the reconstruction quality as it favours the learning of different aspects of the reflectance function.
Rainer~\etal~\cite{gilles2019} use a stereographic projection of the light and view directions in euclidian coordinates as network parameters.
While this parametrisation lends itself well to the modelling of effects like anisotropy, inter-shadowing and masking, which dominate the appearance of sparsely sampled spatially-varying materials, it is not well-suited to reconstruct specular highlights (as can be seen in Figure~\ref{fig:btfnet}), which are much more noticeable in densely sampled uniform materials.
In contrast, we use the Cartesian vectors $\mbf{h}$ and $\mbf{d}$ of the Rusinkiewicz parameterisation~\cite{rusinkiewicz1998} for directions, which are a much better suited set of variables to encode specular lobes.

During training we compute the difference between predicted and
ground-truth BRDF data using a logarithmic loss applied to cosine
weighted reflectance values:
\begin{align}
    \text{Loss} = \bigl|\; \log(1 + f_r^\text{true}\cos\theta_i) - \log(1 + f_r^\text{pred}\cos\theta_i) \,\bigr|\;,
\end{align}

Conveniently, the architecture allows for unstructured sampling of the angular domain, allowing for a BRDF-aware adaptive random sampling of the upper hemisphere, for a total of $8\times10^5$ samples. We draw random uniform samples
of the Rusinkiewicz parameterisation angles,
which emphasises directions close to the specular highlight. In
Section~\ref{sec:res:reconstruction} we show that this is critical for
accurate encoding of the specular highlights. The loss stabilises
after $5$ epochs for the more diffuse materials in
Matusik et al.'s MERL database~\cite{mpbm03} (detailed in
Section~\ref{sec:met:merl})
while the most mirror-like ones can take up-to $90$ epochs
(between $10$ seconds and $3$ minutes on GPU).

NBRDF networks can be used to encode
both isotropic and anisotropic materials. The latter introduce a
further dependence on the Ruinskiewicz angle $\phi_h$, which must be learnt by
the network. Following our sampling strategy, during training we
draw random uniform samples from all four Rusinkiewicz angles,
increasing the total number of samples 5-fold to compensate for
the increased complexity of the BRDF functional shape.
In Section~\ref{sec:res:anisotropic} we analyse the reconstruction of
anisotropic materials from the RGL database~\cite{rgl}, which contains
$51$ isotropic and $11$ anisotropic measured materials.

\subsection{NBRDF autoencoder}
\label{sec:met:ae}

\begin{figure}[htb] 
  \centering
  \includegraphics[width=0.8\linewidth]{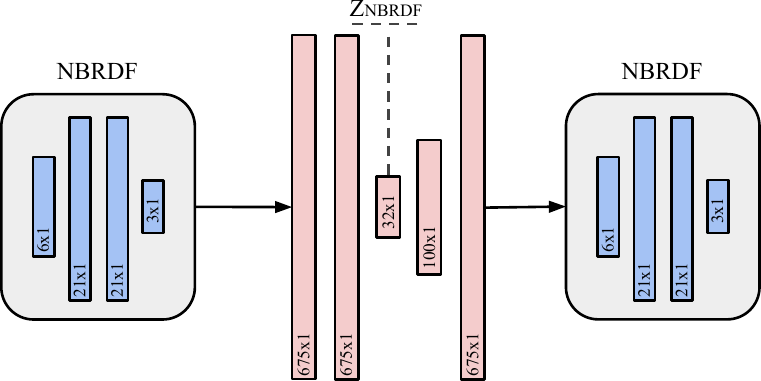}
  \caption{\label{fig:nbrdf-ae}%
    \textit{NBRDF autoencoder}. Input and output
    are NBRDF networks of shape $6\times21\times21\times3$, flattened
    to 1D vectors of $675$ values.}
\end{figure}

Figure~\ref{fig:nbrdf-ae} shows our architecture for an autoencoder
that learns a latent representation for NBRDFs. Input and output are
the flattened weights of an NBRDF, which are further compressed by the network
into short embeddings. In effect, the autoencoder learns to predict the
weights of an NBRDF neural network. We typically use NBRDF encodings
with two hidden layers $6\times21\times21\times3$ for a total of $675$
parameters and encode them into embeddings of $32$ values.

In addition to further compressing the NBRDF representations, the
autoencoder provides consistent encodings of the MERL materials that
can be interpolated to generate new materials, as demonstrated in
Section~\ref{sec:res:newmaterials}. Additionally, we show in
Sections~\ref{sec:met:sampling} and \ref{sec:res:sampling} that these
consistent encodings can be used to predict parameters that can be
leveraged for importance sampling.

Training of the autoencoder is performed using NBRDFs pre-trained with
materials from MERL, employing a 80\%-20\% split between training and
testing materials.
To compensate for the limited availability of measured materials,
we augment our data by applying all permutations of RGB channels for
each material in the training set.
The training loss used is image-based: our custom
loss layer uses the predicted $675\times1$ vector to construct an
NBRDF network of the original shape ($6\times21\times21\times3$),
and evaluates it to produce small renderings ($64\times64$) of a
sphere illuminated by
a non-frontal
directional light
with $\theta_l = 45^\circ$, previously reported to produce
more accurate results than headlight illumination on image-based
BRDF fittings~\cite{sztrajman2019}.
A fixed tone mapping (simple gamma curve with
$\gamma=2.2$ and low values bottom-clamped to 
$10^{-12}$) is then
applied to the sphere renderings, and the loss is computed as
point-by-point MSE. The loss computation involves a differential
implementation of the rendering pipeline for direct illumination and
subsequent tone mapping, in order to keep the computation
back-propagatable.
Notably, applying a more traditional, non-image-based loss that attempts to match the input NBRDF weights fails to reconstruct the original appearances of the encoded materials.

\subsection{Importance Sampling}
\label{sec:met:sampling}

Importance sampling of BRDFs requires producing angular samples with a
probability density function (PDF) approximately proportional to the
BRDF. This can be accomplished by computing the inverse cumulative
distribution function (inverse CDF) of the PDF, which constitutes
a mapping between a uniform distribution and the target distribution.
The computation of the inverse CDF of a PDF usually requires costly
numerical integrations; however, for a set of parametric BRDF models,
such as Blinn-Phong or GGX, this can be done analytically.

\begin{figure}[htpb] 
  \centering
  \includegraphics[width=0.85\linewidth]{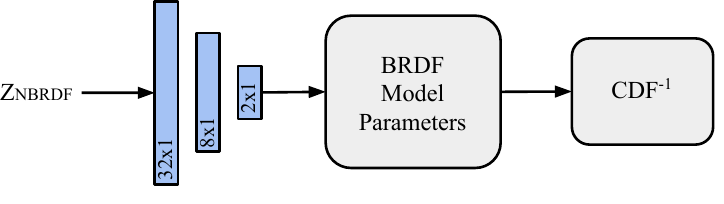}
  \caption{\label{fig:brdf2phong}%
  Scheme for computation of inverse CDF from an NBRDF: we train a network to map from latent NBRDF embeddings to importance sampling parameters of a chosen analytic BRDF model.}
\end{figure}

Our proposed method for quick inverse CDF computation is based on a shallow neural
network, shown in Figure~\ref{fig:brdf2phong}, that learns the mapping between the
embeddings generated by the NBRDF autoencoder and a set of model parameters from an
invertible analytic BRDF. In essence, the network learns to fit NBRDFs to an analytic
model, an operation that is commonly performed through nonlinear optimisation, which
is comparatively slow and prone to get lodged in local minima.

We use Blinn-Phong as target model for our prediction. Although it contains a total
of $7$ model parameters, its associated PDF is monochrome and can be defined by
only $2$ parameters, associated with the roughness of the material and the relative
weight between specular and diffuse components. Hence, we train our network to learn
the mapping between the NBRDF's $32$-value embeddings and the Blinn-Phong importance
sampling parameters. Although the predicted PDF is an approximation of the original
NBRDF, the resulting sampling is unbiased due to the exact correspondence between
the sampling PDF and its inverse CDF, as shown in Section~\ref{sec:res:sampling}. 

\subsection{MERL Database}
\label{sec:met:merl}

The MERL BRDF database~\cite{mpbm03} contains reflectance measurements
from $100$ real-world materials, with a dense sampling of
directions given directly in terms of the spherical angles ($\theta$,
$\phi$) of the $\mbf{h}$ and $\mbf{d}$ vectors from the Rusinkiewicz
parameterisation~\cite{rusinkiewicz1998}:
\setlist[description]{labelindent=0em,labelwidth=2em,labelsep=0em,leftmargin=2em,itemsep=0.667ex}
\begin{description}
\item[$\theta_h$:]$90$ samples from $0$ to $90$, with inverse square-root sampling that emphasises low angles.
\item[$\theta_d$:]$90$ uniform samples from $0$ to $90$.
\item[$\phi_d$:]$180$ uniform samples from $0$ to $180$. Values from $180$ to $360$ are computed by applying Helmholtz reciprocity.
\end{description}
Isotropic BRDFs are invariant in $\phi_h$, so the MERL database, which
was created using a measurement setup relying on isotropic
reflectance~\cite{marschner1999imagebased}, omits $\phi_h$. Counting
all samples for the three colour channels, each material in MERL is
encoded in tabular format with $4.4\times10^6$ reflectance values (approx. 34 MB).

\section{Results}
\label{sec:res}
\begin{figure}[H] 
  \centering
  \includegraphics[width=\linewidth]{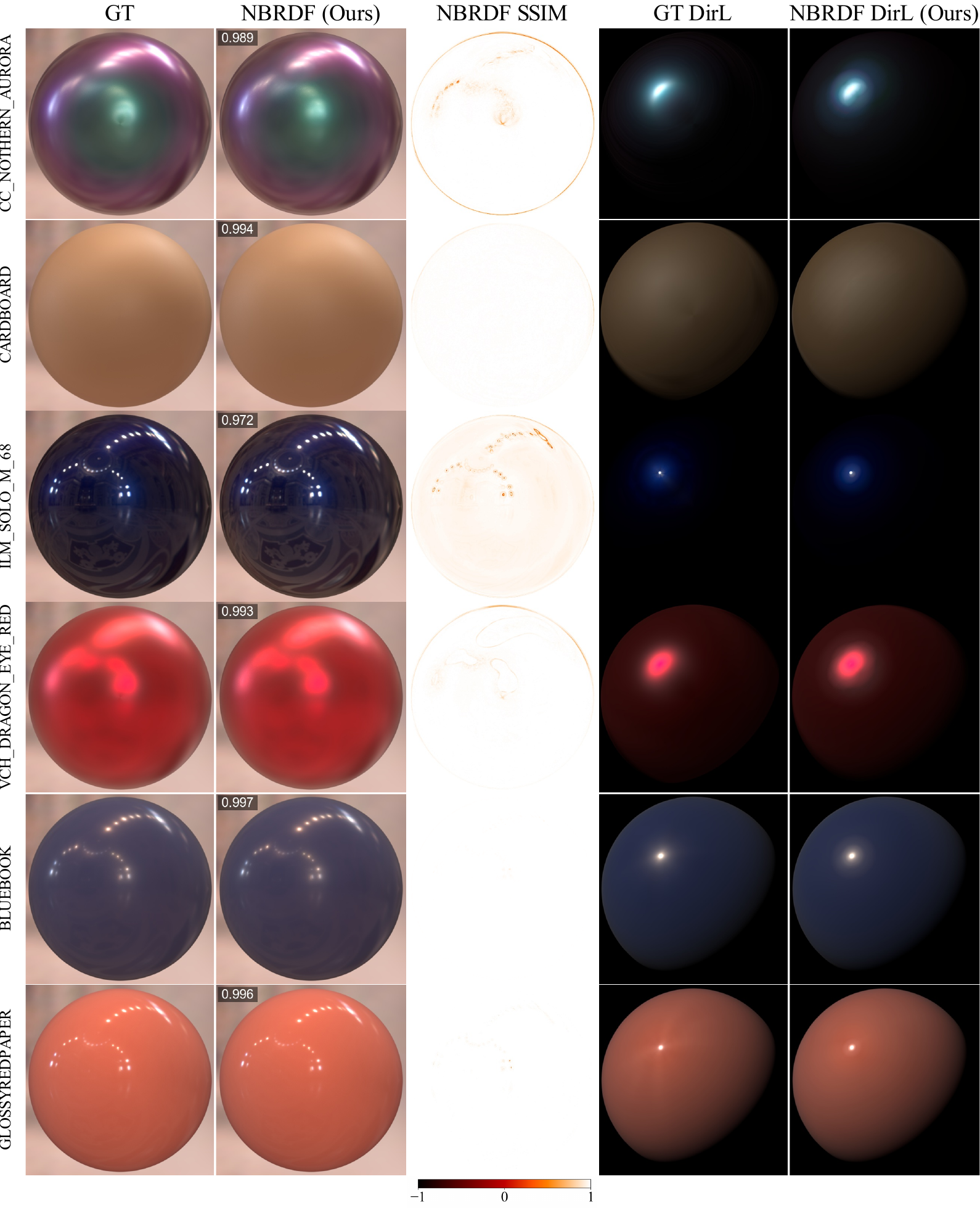}
  \caption{\label{fig:nielsen}%
  NBRDF reconstruction of measured materials from the RGL database~\cite{rgl} (top 4) and from Nielsen~\etal~\cite{nielsen2015} (bottom 2), using environment map and directional illuminations.}
\end{figure}

In this section, we analyse our results on the reconstruction and importance sampling
of measured materials. Although we centre most of our analysis on materials from the MERL database,
we show that our approach can be applied to any source of measured BRDFs, as displayed
in Figure~\ref{fig:nielsen}.
Reconstruction results for the complete
set from MERL~\cite{mpbm03} and RGL~\cite{rgl} databases can be found in the supplemental material.
In addition, we have included our implementation of the NBRDF training in Keras~\cite{keras}, a Mitsuba plugin to render using our representation, and a dataset of pre-trained NBRDFs for materials from the MERL~\cite{mpbm03}, RGL~\cite{rgl} and Nielsen~\etal~\cite{nielsen2015} databases.

\begin{figure*}[htb] 
  \vspace{0.5em}
  \centering
  \includegraphics[width=\linewidth]{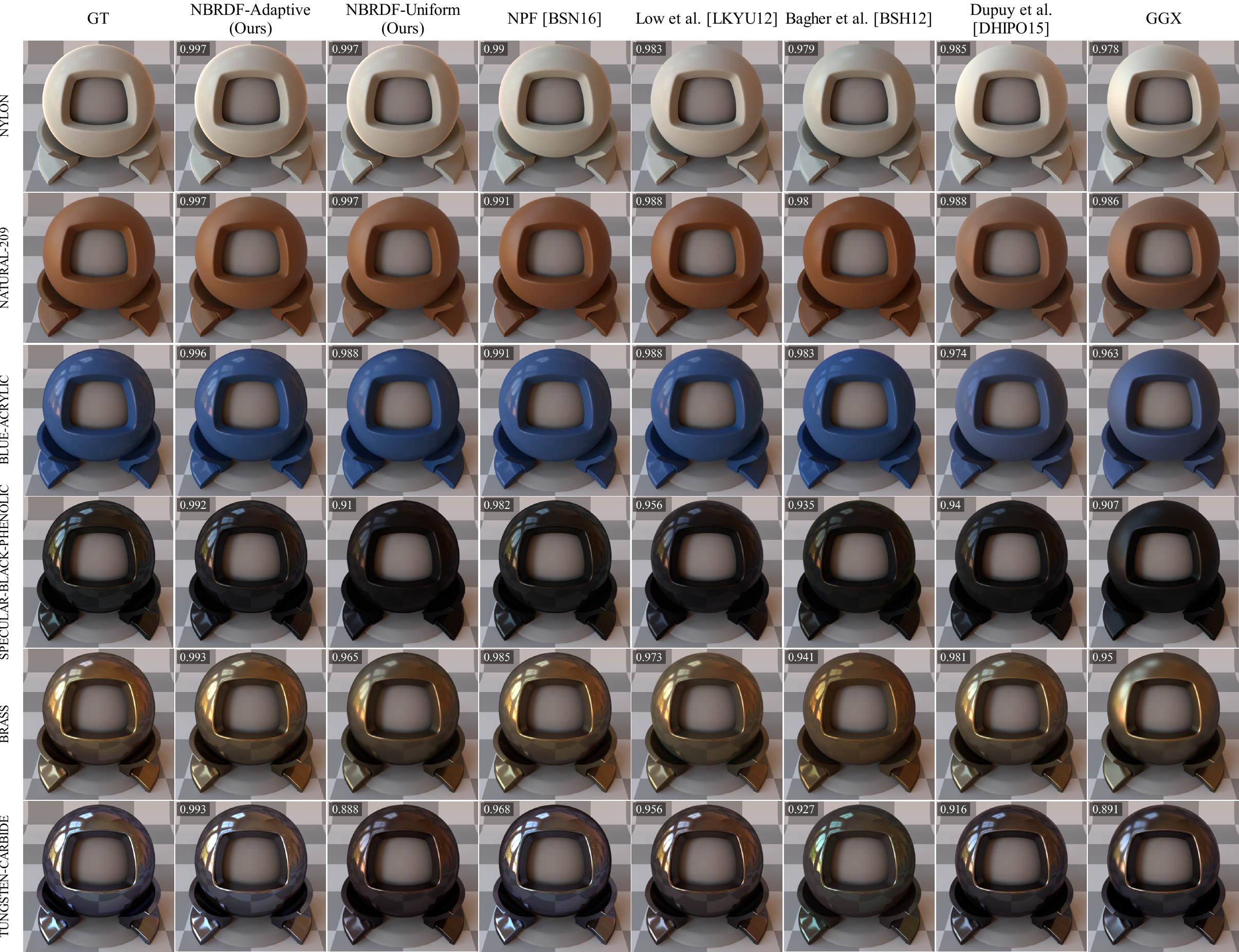}
  \caption{\label{fig:reconstruction}%
  Reconstruction of MERL materials using different BRDF representations, including the average SSIM value for each image. For both NBRDF columns we utilised a fixed network size of $675$ weights ($6\times21\times21\times3$).}
\end{figure*}%

\begin{figure*}[htb] 
  \centering
  \vspace{0.5em}
  \includegraphics[width=\linewidth]{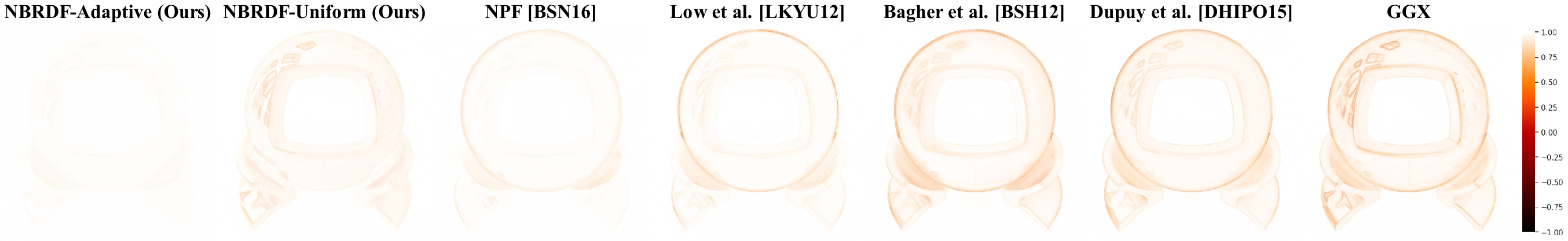}
  \caption{\label{fig:average}%
  Average SSIM over all MERL materials for different BRDF representations.}
\end{figure*}

\subsection{BRDF Reconstruction}
\label{sec:res:reconstruction}

Figure~\ref{fig:reconstruction} shows reconstruction performance on a visually diverse set of materials of the MERL database,
for different approaches.
We qualitatively compare the methods through renderings of a scene with environment map illumination. 
Ground truth is produced by interpolating the tabulated MERL data.
The comparison reveals that most methods struggle with one particular type of materials:
a GGX fit tends to blur the highlights, Bagher~\etal~\cite{bagher2012} on the other hand achieve accurate specular highlights, but the diffuse albedo seems too low overall.
Out of all the proposed representations, our method produces the closest visual fits, followed by
NPF~\cite{bagher2016}, a non-parametric BRDF fitting algorithm recently cited as state-of-the-art~\cite{rgl}.
As detailed in Section~\ref{sec:rel:compression}, a recent data-driven BRDF model by Dupuy and Jakob~\cite{rgl} also compared favourably against NPF, although at an increased storage requirement.

\begin{table*}[htb] 
\centering
\begin{tabular}{@{}lccccccl@{}}
\toprule
                                     & MAE                          & RMSE                         & SSIM \\
\midrule
NBRDF Adaptive Sampling              & $\mathbf{0.0028 \pm 0.0034}$ & $\mathbf{0.0033 \pm 0.0038}$ & $\mathbf{0.995 \pm 0.008}$ \\
NBRDF Uniform Sampling               & $0.0072 \pm 0.0129$          & $0.0078 \pm 0.0134$          & $0.984 \pm 0.029$ \\
NPF~\cite{bagher2016}                & $0.0056 \pm 0.0046$          & $0.0062 \pm 0.0047$          & $0.990 \pm 0.008$ \\
Low~\etal~\cite{low2012} (ABC)       & $0.0080 \pm 0.0070$          & $0.0088 \pm 0.0075$          & $0.986 \pm 0.012$ \\
Bagher~\etal~\cite{bagher2012} (SGD) & $0.0157 \pm 0.0137$          & $0.0169 \pm 0.0145$          & $0.974 \pm 0.027$ \\
Dupuy~\etal~\cite{dupuy2015}         & $0.0174 \pm 0.0143$          & $0.0190 \pm 0.0151$          & $0.976 \pm 0.021$ \\
GGX                                  & $0.0189 \pm 0.0118$          & $0.0206 \pm 0.0126$          & $0.969 \pm 0.024$ \\
\bottomrule
\end{tabular}
\caption{\label{table:reconstruction}%
  Average image-based losses of representation methods from Figure~\ref{fig:reconstruction} over all MERL materials.}
\end{table*}

\begin{figure}[htb] 
  \centering
  \includegraphics[width=0.9\linewidth]{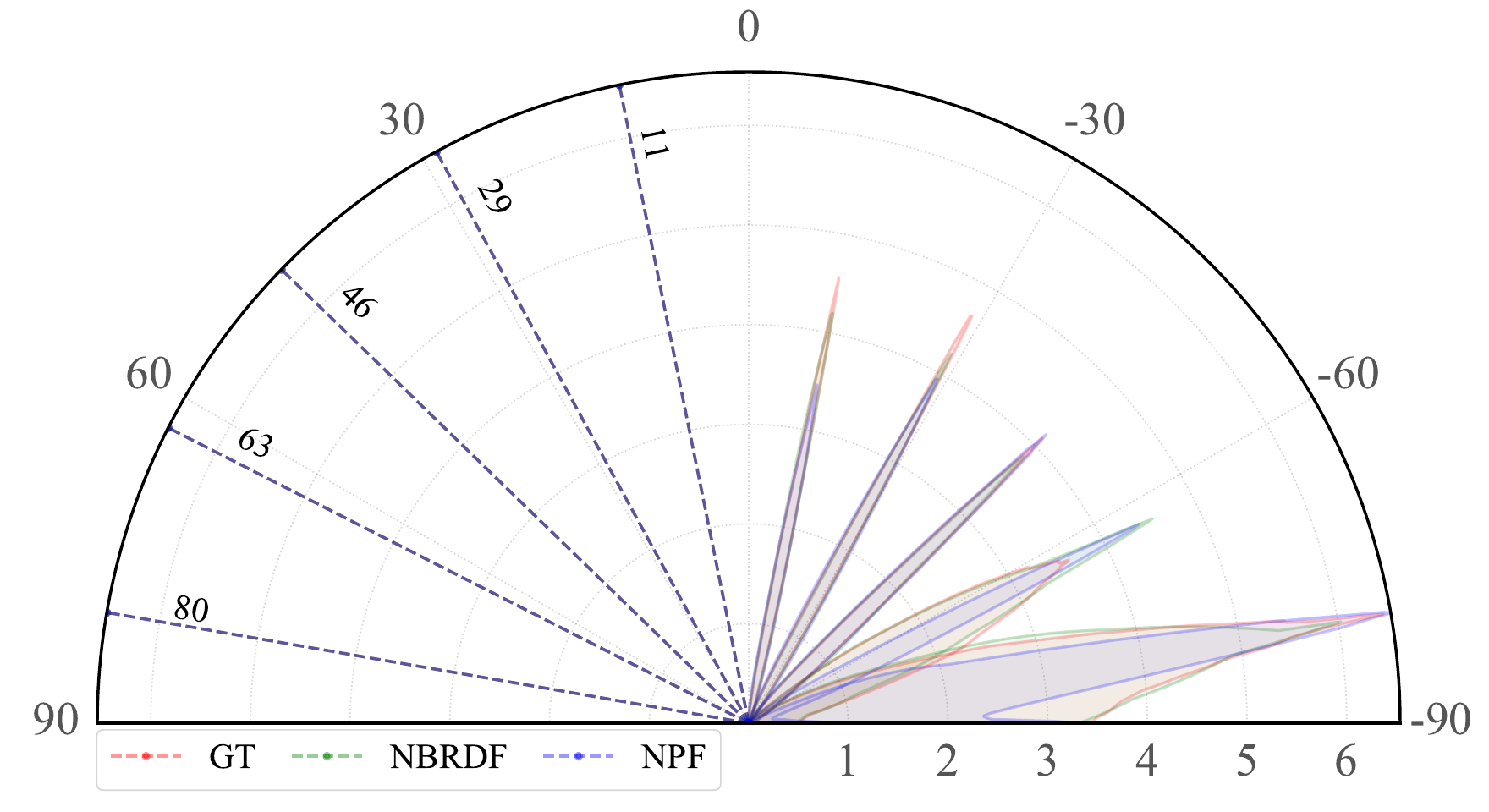} \\
  \includegraphics[width=0.9\linewidth]{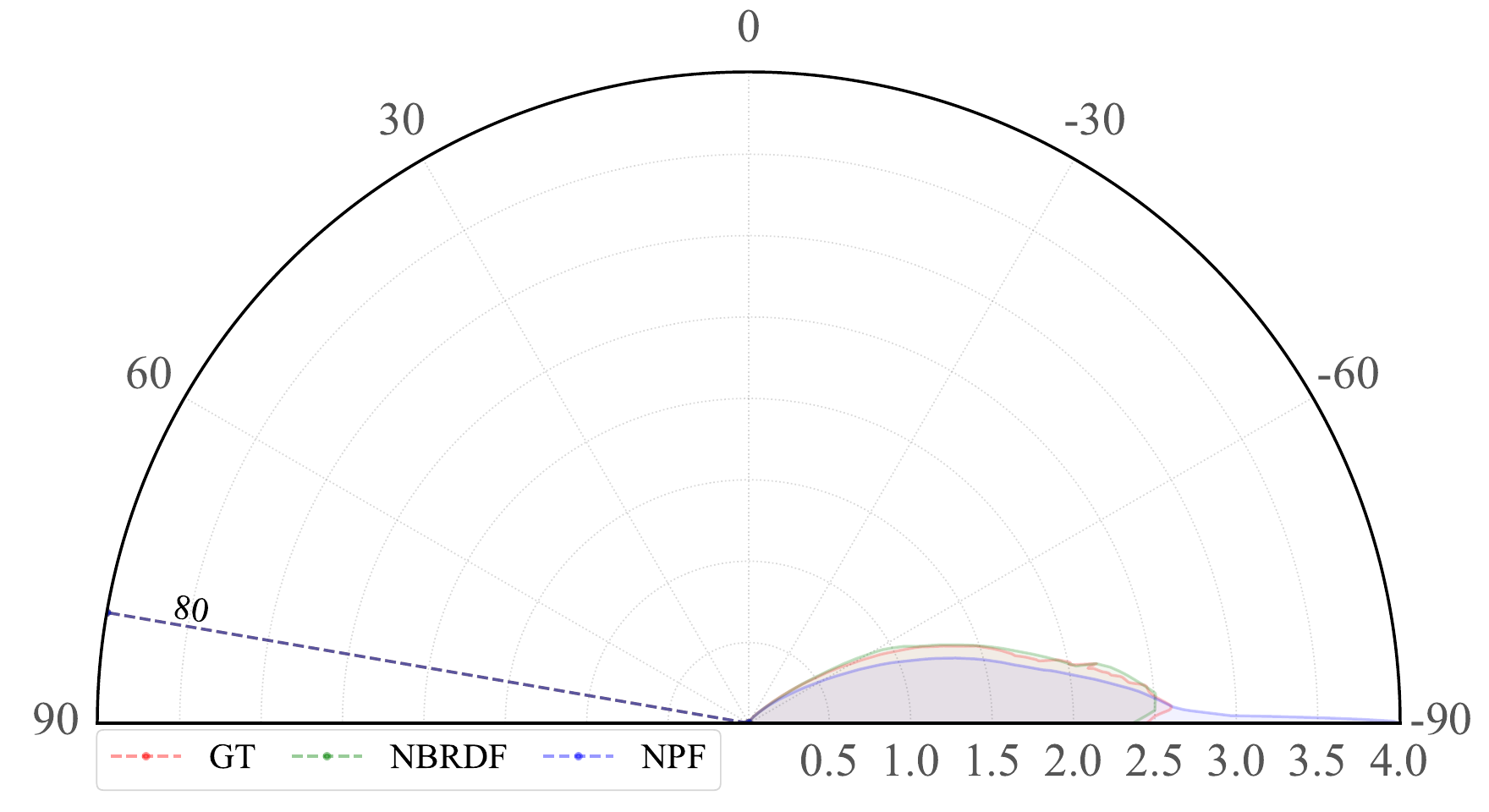}
  \caption{\label{fig:polar}%
  Polar plots (log scale) comparing NPF~\cite{bagher2016} and our NBRDF fittings with ground truth lobes for fixed incident inclination angles $\theta_i$. \emph{Top}: \texttt{grease-covered-steel}. \emph{Bottom}: \texttt{black-oxidized-steel} with a single fixed $\theta_i$ at $80^o$.}
\end{figure}

A quantitative analysis of the results, seen in Figure~\ref{fig:average} and Table~\ref{table:reconstruction}, shows that our representation outperforms the other methods in multiple image-based error metrics.
In particular, NPF~\cite{bagher2016} seems to lose fitting accuracy at very grazing angles, which is where the error is the highest on average (see Figure~\ref{fig:average}). A more detailed analysis of the functional shape of the NPF lobes confirms this
observation. In Figure~\ref{fig:polar} we display polar plots (in log scale) of the specular lobes of two materials from MERL, comparing NBRDF and NPF fittings with ground truth for fixed incident angles. For low values of incident inclination $\theta_i$ there is generally good agreement between all representations, while for grazing angles only NBRDFs are able to match the original shape. Furthermore,
in the bottom plot we observe that NPF tends to produce unusually long tails. In the supplemental material we provide polar plot comparisons for the the full set of MERL materials.

\begin{figure}[htb]
  \centering
  \includegraphics[width=.9\linewidth]{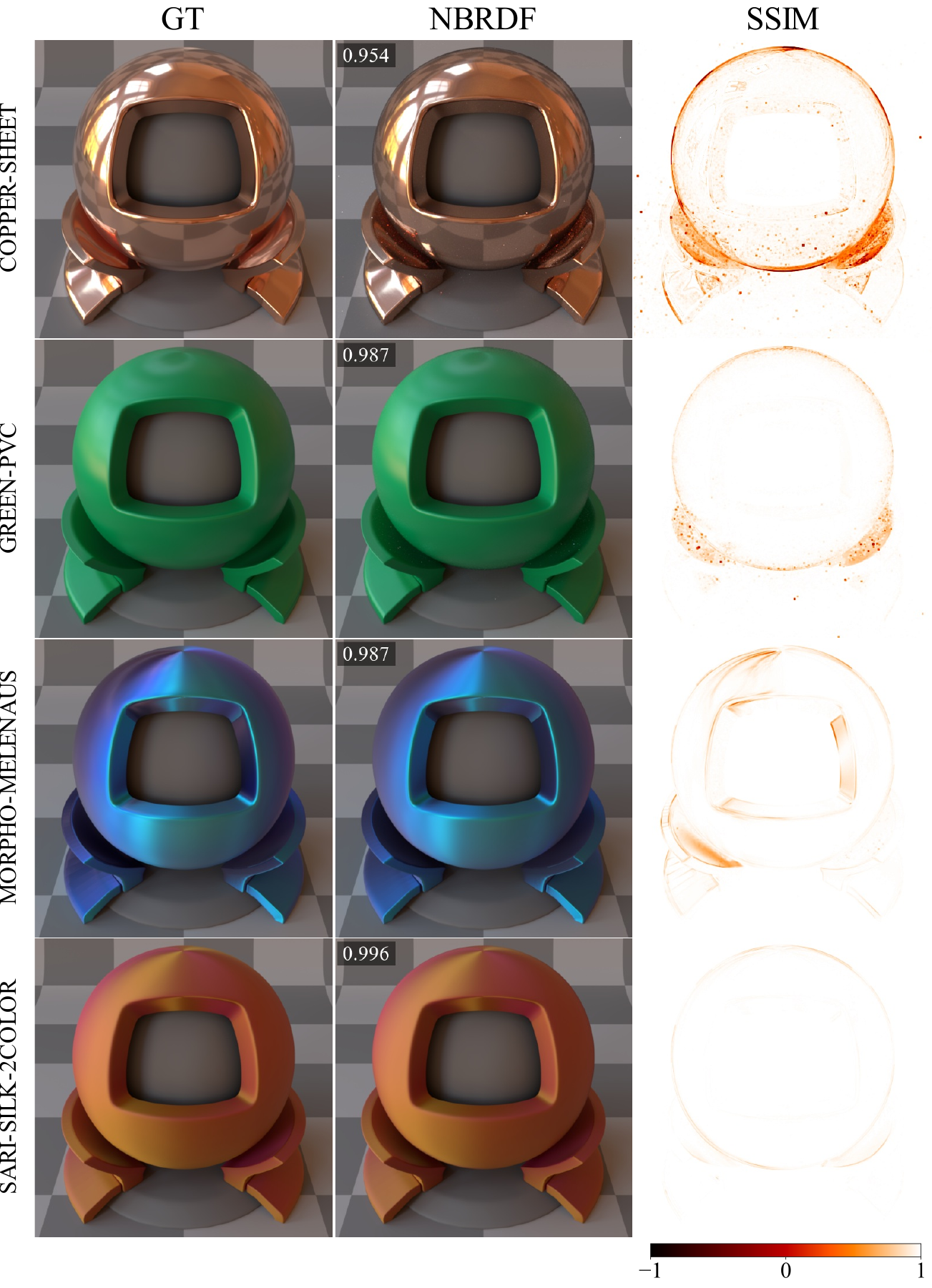}
  \caption{\label{fig:aniso_results}%
  Neural BRDF reconstruction of anisotropic materials from the RGL database~\cite{rgl}.}
\end{figure}%

\begin{figure*}[htb]
  \centering
  \includegraphics[width=\linewidth]{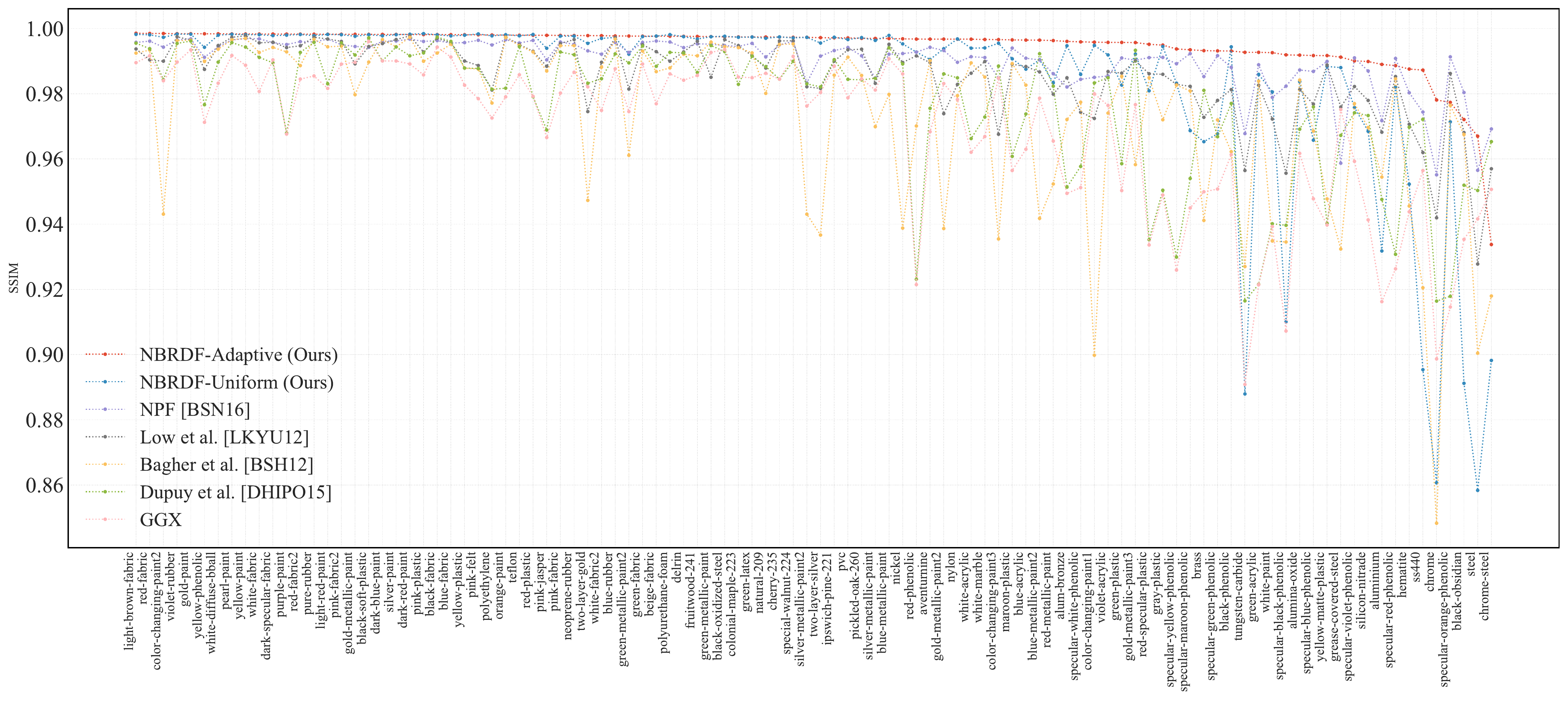}
  \caption{\label{fig:ssimplot}
  SSIM error for all materials from the MERL database using the BRDF reconstruction methods from Figure~\ref{fig:reconstruction}.}
\end{figure*}%

One of the key components in successfully training the NBRDF networks is the angular sampling of the training loss.
If training samples are concentrated near the specular lobe, the NBRDF will accurately reproduce the highlights.
On the other hand, if the samples are regularly distributed, the lambertian reflectance component will be captured more efficiently.
We hence employ a BRDF-aware adaptive sampling of angles during training that emphasises samples close to the reflectance lobes.
In practice, we uniformly (randomly) sample the spherical angles of the Rusinkiewicz parameterisation ($\theta_h$, $\theta_d$ and $\phi_d$),
which results in a sample concentration around the specular direction, while retaining sufficient coverage of the full hemisphere.
Table~\ref{table:reconstruction} shows that this adaptive strategy for training sample generation produces much better results over the whole database and allows us to outperform analytic model fits in various error metrics.

Finally, in Figure~\ref{fig:ssimplot} we display the SSIM error for all materials from the MERL database, and for all discussed reconstruction methods. Our NBRDF adaptive-sampling outperforms other methods for almost all materials, with the exception of a small number of highly specular materials. Please refer to the supplemental material for a full detail of reconstructions,
including all materials from the MERL and RGL~\cite{rgl} databases.

\subsection{Reconstruction of Anisotropic Materials}
\label{sec:res:anisotropic}

In Figure~\ref{fig:aniso_results} we display the NBRDF reconstructions of multiple
anisotropic materials from the RGL database~\cite{rgl}. The networks used are the same
as shown in the isotropic results of Figure~\ref{fig:reconstruction}
(\ie $6\times21\times21\times3$ for a total of $675$ weights).
The reconstruction of the anisotropy is surprisingly robust, especially taking into
account the compactness of the network size. There are, however, more perceivable differences
in the visual fits than in the NBRDF isotropic encodings, which is reflected on the
average SSIM error: $0.981 \pm 0.016$.
Lower reconstruction errors can be achieved by increasing the network size of the encoding
NBRDF, providing great control over the level-of-detail of the representation.
In Section~\ref{sec:res:performance} we will analyse the dependence of the reconstruction 
error with the network size, comparing with other representations in terms of memory footprint.

\subsection{Latent Space of Materials}
\label{sec:res:newmaterials}

\begin{figure}[htb]
  \centering
  \includegraphics[width=\linewidth]{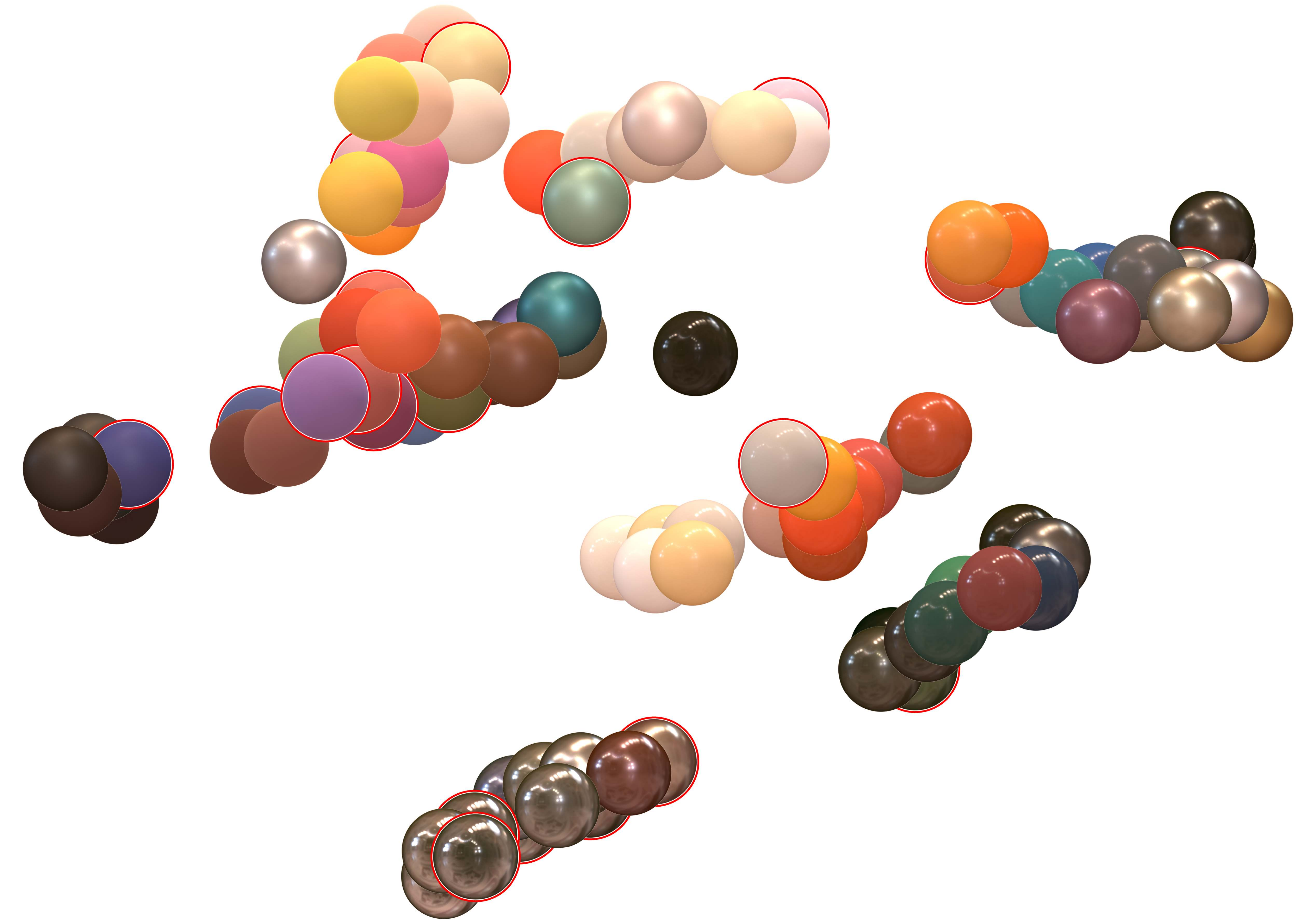}
  \caption{\label{fig:tsne}%
    $t$-SNE clustering of MERL latent embeddings produced by the NBRDF autoencoder.
    Test set materials are indicated in red.
  }
\end{figure}

\begin{figure*}[htb]
  \centering
  \includegraphics[width=\linewidth]{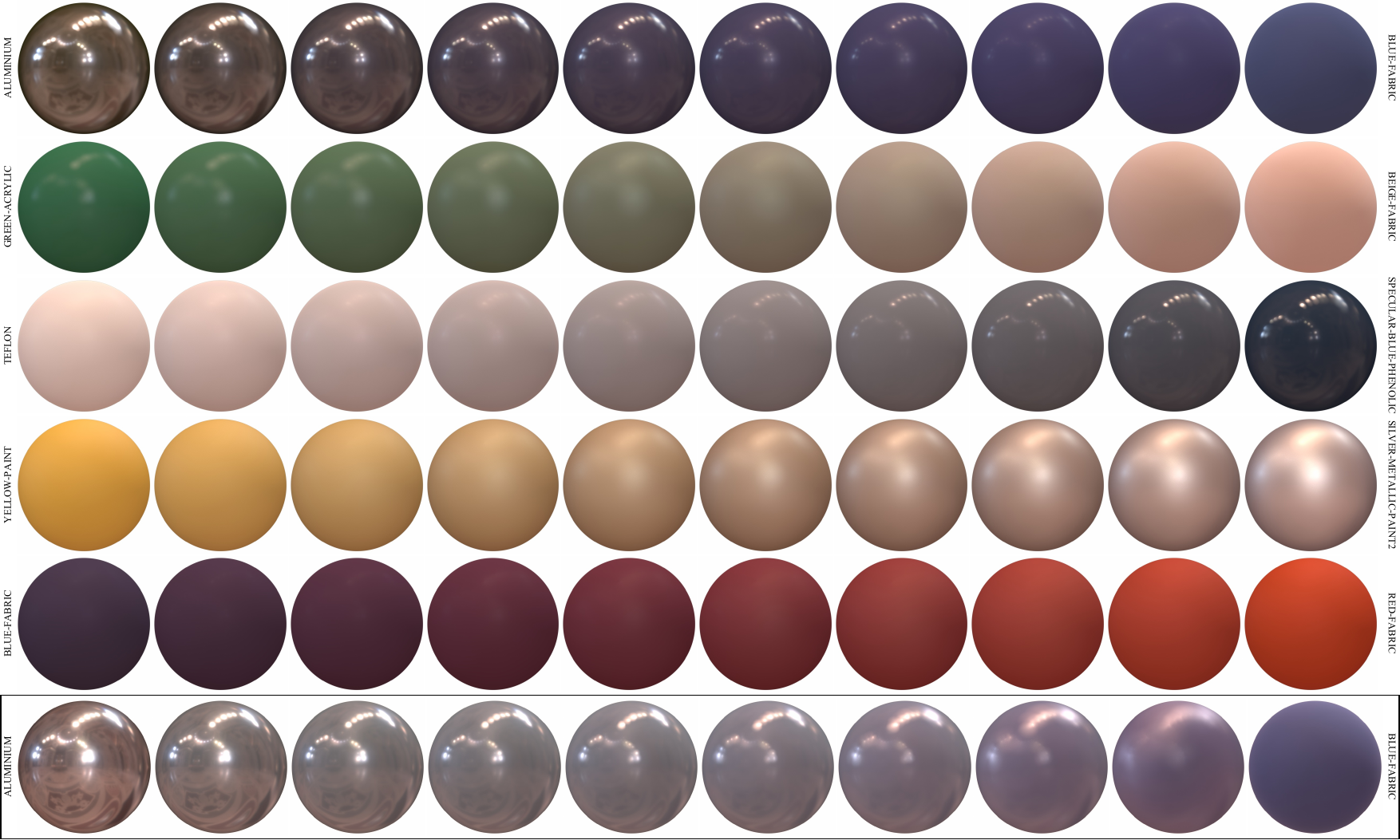}
  \caption{\label{fig:interpolated}%
New materials generated by latent interpolating of MERL BRDFs~\cite{mpbm03}.
Materials on the sides correspond to reconstructed original materials, while the materials in-between were created by uniform interpolation of the original embedding positions in the autoencoder's latent space. \emph{Bottom:} Irregular transition between two materials from the MERL database~\cite{mpbm03}, generated by the direct interpolation of the $675$ parameters of two NBRDF networks.}
\end{figure*}

The generation of a unified encoding of the space of materials
opens up many new possibilities.
We use the NBRDF encodings of MERL materials to train our autoencoder that compresses NBRDFs to a 32-dimensional latent space.

In Table~\ref{table:pca} we summarise various reconstruction error metrics comparing our autoencoding with PCA factorisation
across MERL. Our implementation of PCA follows Nielsen~\etal's~\cite{nielsen2015}, who proposed various improvements over
traditional PCA, most importantly a log-mapping of reflectance values relative to a median BRDF measured over the training set.
The training of both methods was performed with the same 80\%-20\% split of materials from MERL.
The full set of renderings and errors can be found in the supplemental material.
%
\begin{table}[htb] 
\centering
\small
\begin{tabular}{@{}lccccccl@{}}
\toprule
                        & MAE                         & RMSE                        & SSIM \\
\midrule
NBRDF AE                & $\mathbf{0.0178 \pm 0.013}$ & $\mathbf{0.0194 \pm 0.014}$ & $0.968 \pm 0.031$ \\
PCA~\cite{nielsen2015}  & $0.0199 \pm 0.008$          & $0.0227 \pm 0.009$          & $\mathbf{0.982 \pm 0.007}$ \\
\bottomrule
\end{tabular}
\caption{\label{table:pca}%
Average image-based reconstruction losses over all MERL materials for our NBRDF autoencoder and 32-dimensional PCA.}
\end{table}

It is worth noting that the further reduction of NBRDFs from $675$ parameters to $32$ does not necessarily result
in an effective compression of the representation in a practical use case, since the memory footprint
of the decoder is roughly equivalent to $105$ NBRDFs. In addition, preserving the maximum reconstruction quality requires storing the original NBRDF, since the autoencoder reduction
inevitably leads to a
degradation of the appearance after the decoding;
however, this is not an issue as the main application of the autoencoder lies in the material embedding.
Figure~\ref{fig:tsne} shows a $t$-SNE clustering of the latent embedding learned by the autoencoder.
The projection to the latent space behaves sensibly, as materials with similar albedo or shininess cluster together.
This 32-dimensional encoding is the basis for our subsequent importance sampling parameter prediction.

The stability of the latent space is further demonstrated in Figure~\ref{fig:interpolated}, where we linearly interpolate, in latent space, between encodings of MERL materials, and visualise the resulting decoded materials.
In contrast, the bottom row of Figure~\ref{fig:interpolated} shows the direct interpolation of the $675$ parameters from two individually-trained NBRDF networks. Noticeably, this does not lead to a smooth transition of the specular properties of the two materials.

\subsection{Importance Sampling}
\label{sec:res:sampling}

\begin{figure*} 
   \centering
   \includegraphics[width=0.9\linewidth]{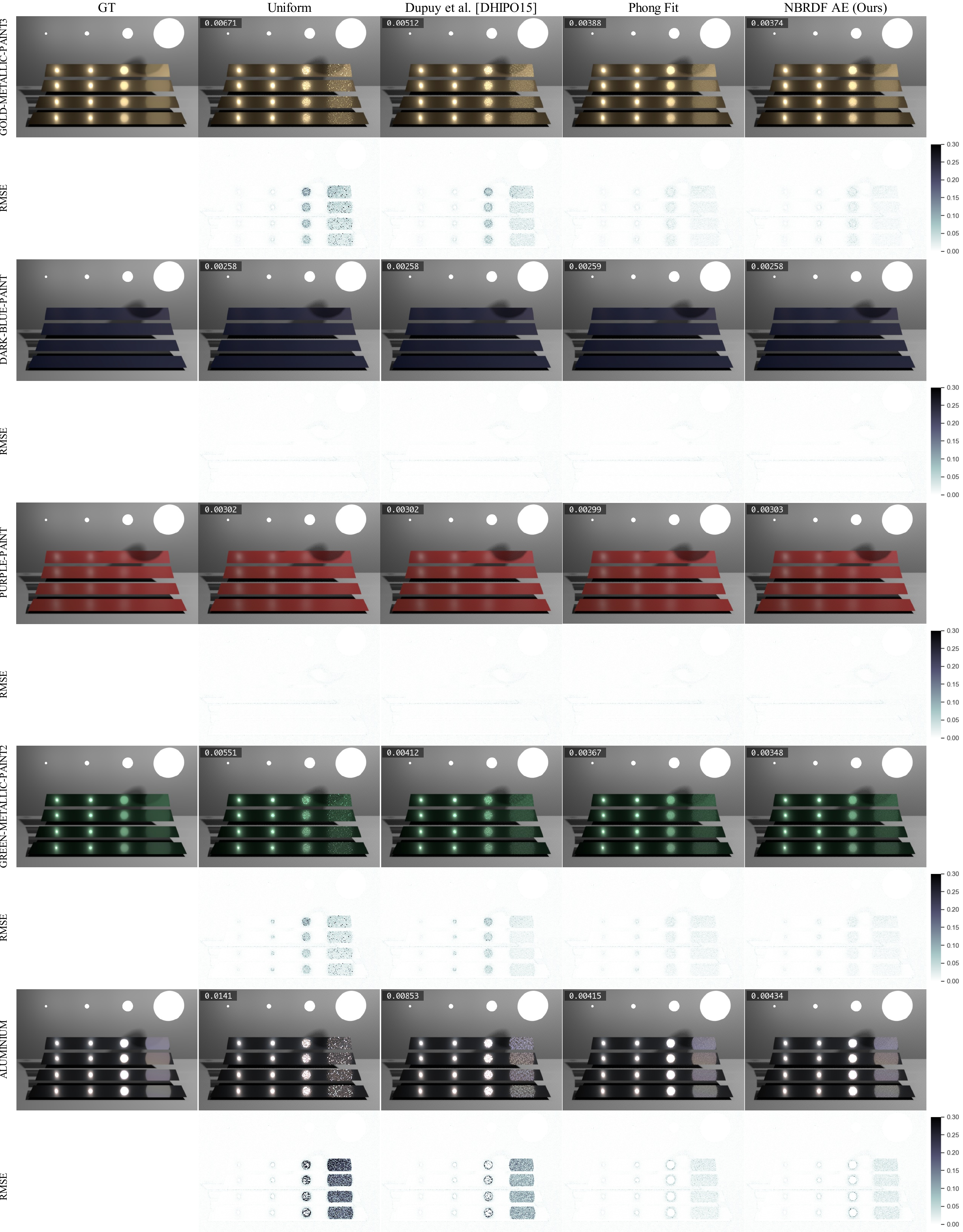}
   \caption{\label{fig:isampling}%
   Importance sampling of testset materials from the MERL database, using $64$ samples per pixel. \emph{Left to right:} Ground Truth ($6400$ spp), Uniform sampling, Phong sampling (Optimised), Phong sampling (Our). Bottom: RMS errors.}
\end{figure*}

We leverage the stable embedding of materials provided by the autoencoder to predict importance sampling parameters.
In practice, we train a network to predict the 2 Blinn-Phong distribution parameters that are used in the importance sampling routine.
We train on a subset of materials from the MERL database, using fitted Blinn-Phong parameters from Ngan~\etal~\cite{ngan2005} as labels for supervised training.
In Figure~\ref{fig:isampling} we compare and analyse the effect of different importance
sampling methods, applied to multiple materials from MERL unseen by our importance sampling prediction
network.
Renderings are produced with $64$ samples per pixel, with the exception of the ground truth at $6400$ spp.
Each column is associated with a different importance sampling method, with all reflectance
values begin evaluated from the original tabulated MERL data.
We compare uniform sampling, Blinn-Phong distribution importance sampling (with optimised parameters, and predicted parameters from our network), and Dupuy~\etal's~\cite{dupuy2015} routine. 
Even though a Blinn-Phong lobe is not expressive enough to accurately describe and fit the captured data, the parameters are sufficient to drive an efficient importance sampling of the reflectance distribution.
Depending on the material, the predicted Blinn-Phong parameters can even reveal themselves better suited for importance sampling than the optimised Blinn-Phong parameters.
%

\begin{figure*} 
   \centering%
   \includegraphics[width=\linewidth]{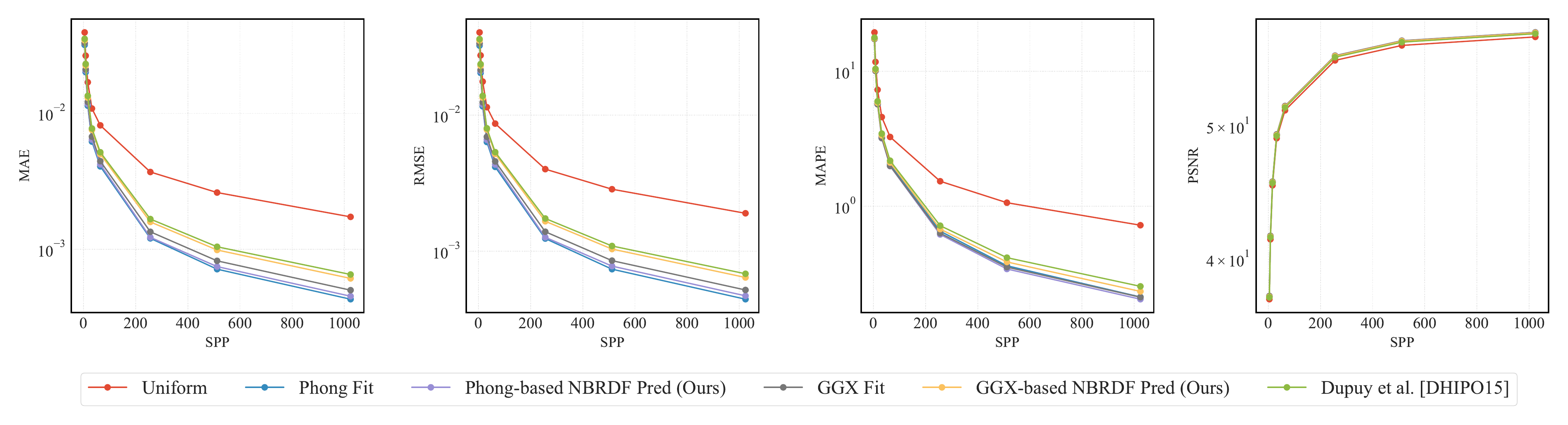}
   \caption{\label{fig:isampling_plots} Average errors (log scale) vs SPP for all 20 MERL testset materials using the Veach scene from Figure~\ref{fig:isampling}. \emph{Left to right:} MAE, RMSE, MAPE, PSNR.}
\end{figure*}

\begin{figure*} 
   \centering%
   \includegraphics[width=\linewidth]{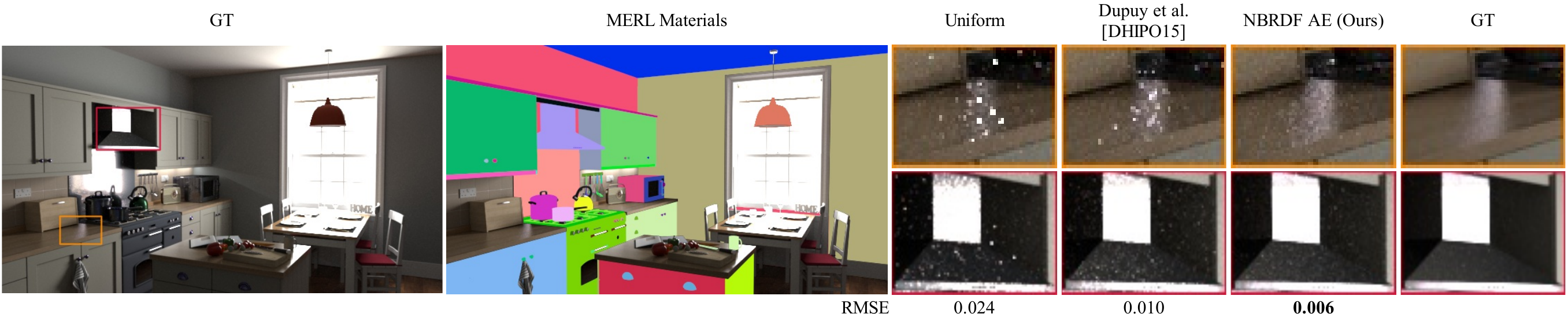}
   \caption{\label{fig:kitchen_renders}%
   Importance sampling of kitchen scene using $6400$ and $64$ samples per pixel respectively for GT and compared methods. Most materials in the scene, shown in flat-colour in the central image, have been replaced by MERL testset materials.}
\end{figure*}

\begin{figure*} 
   \centering
   \includegraphics[width=\linewidth]{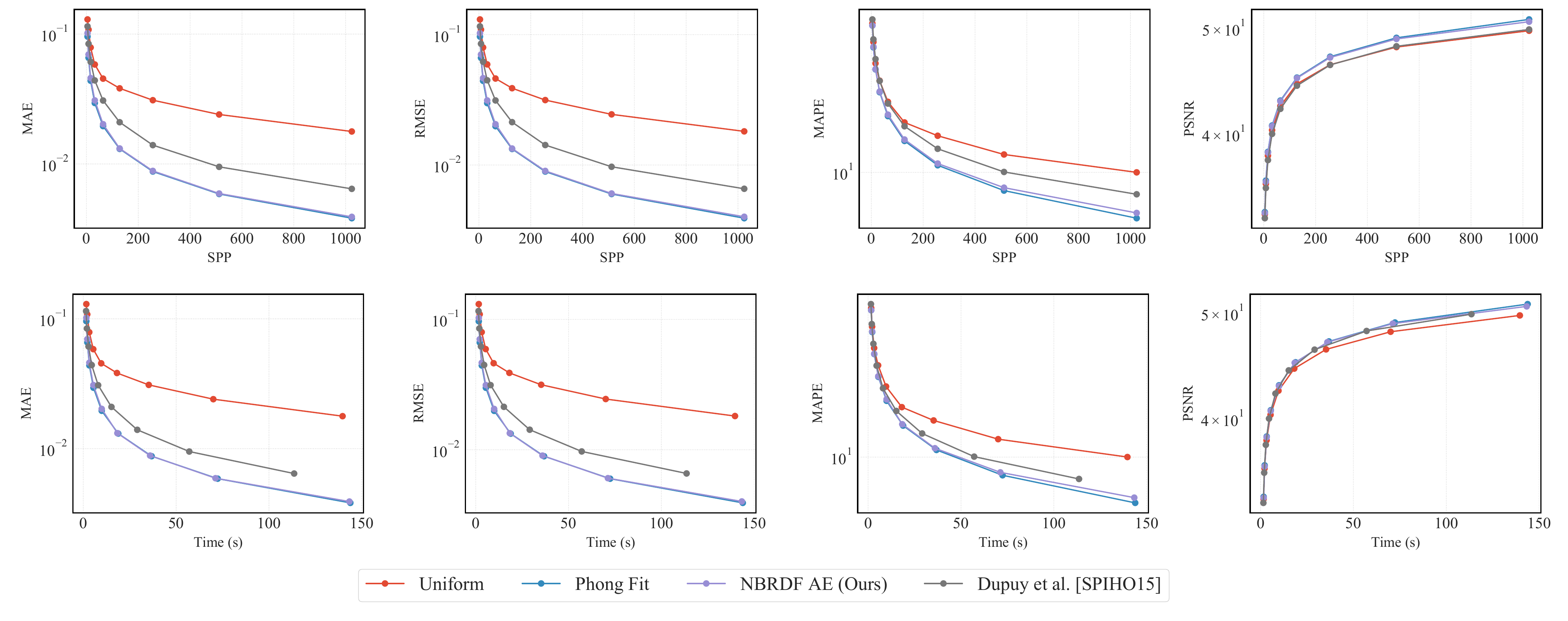}
   \caption{\label{fig:kitchen_metrics} Average errors (log scale) vs SPP (Top) and render time (Bottom) for MERL testset materials in the kitchen scene from Figure~\ref{fig:kitchen_renders}. \emph{Left to right:} MAE, RMSE, MAPE, PSNR.}
\end{figure*}

In addition to this image-based comparison, in Figure~\ref{fig:isampling_plots} we plot multiple error metrics as a function of samples per pixel, to compare the respective sampling methods. 
Both Phong and GGX-driven importance sampling converge quickly and keep a significant lead on uniform sampling.
As shown in the plots, our importance sampling prediction can be tuned to GGX parameters (optimised labels from Bieron and Peers~\cite{bieron2020}) as well as to Blinn-Phong parameters, or any arbitrary distribution.
For simplicity, we choose the Blinn-Phong distribution: more advanced models will provide a better reconstruction, but not necessarily provide a better sampling routine.
More complex models might fit the specular lobe more precisely, but neglect other reflectance components of the data, such as sheen in fabric datasets for instance.

In Figure~\ref{fig:kitchen_renders} we show importance sampling results for a complex scene.
The majority of the original BRDFs in the scene have been replaced by materials from the MERL
database, from the test set of our importance sampling parameter prediction network.
We show crops from the renderings and compare our Phong-based importance sampling performance with uniform sampling and the method by Dupuy~\etal~\cite{dupuy2015}. Our method consistently shows
lower noise in the scene, as also reflected in the numerical errors of Figure~\ref{fig:kitchen_metrics} which show a faster convergence for our method
in terms of samples-per-pixel and render time.

\subsection{Computational Performance}
\label{sec:res:performance}
We compare the performance of our combined pipeline (NBRDF reconstruction, with Phong-based importance sampling), to other compact representations that combine fast BRDF evaluation and built-in importance sampling strategies.
All evaluations were performed with CPU implementations in Mitsuba, running on an Intel Core i9-9900K CPU.
Table~\ref{table:speed} shows that an unoptimised implementation of NBRDFs, combined with Phong importance sampling, although slower than other representations, offers comparable rendering performance, even to simple analytic models such as Cook-Torrance.
%
\begin{table} 
\centering
\begin{tabular}{@{}lcccl@{}}
\toprule
								& Rays/sec $(\times10^6)$ & Memory (KB) \\
\midrule
Bagher et al.~\cite{bagher2012} & $10.64$ & $0.13$ \\
RGL~\cite{rgl}					& $10.66$ & $48.0$ \\
NBRDF + PhongIS (Ours) 			& $12.50$ & $2.70$ \\
Cook-Torrance					& $13.59$ & $0.03$ \\
Dupuy et al.~\cite{dupuy2015} 	& $14.05$ & $2.16$ \\
Low et al.~\cite{low2012} 		& $15.13$ & $0.03$ \\
GGX								& $16.82$ & $0.03$ \\
NPF~\cite{bagher2016}			& --	  & $3.20$ \\
\bottomrule
\end{tabular}
\caption{\label{table:speed} Rays traced per second in Mitsuba~\cite{mitsuba} and memory footprint, for different material representations. The NBRDF numbers correspond to the $675$-weights network ($6\times21\times21\times3$).}
\end{table}

\begin{figure}[h!] 
 \centering
 \includegraphics[width=\linewidth]{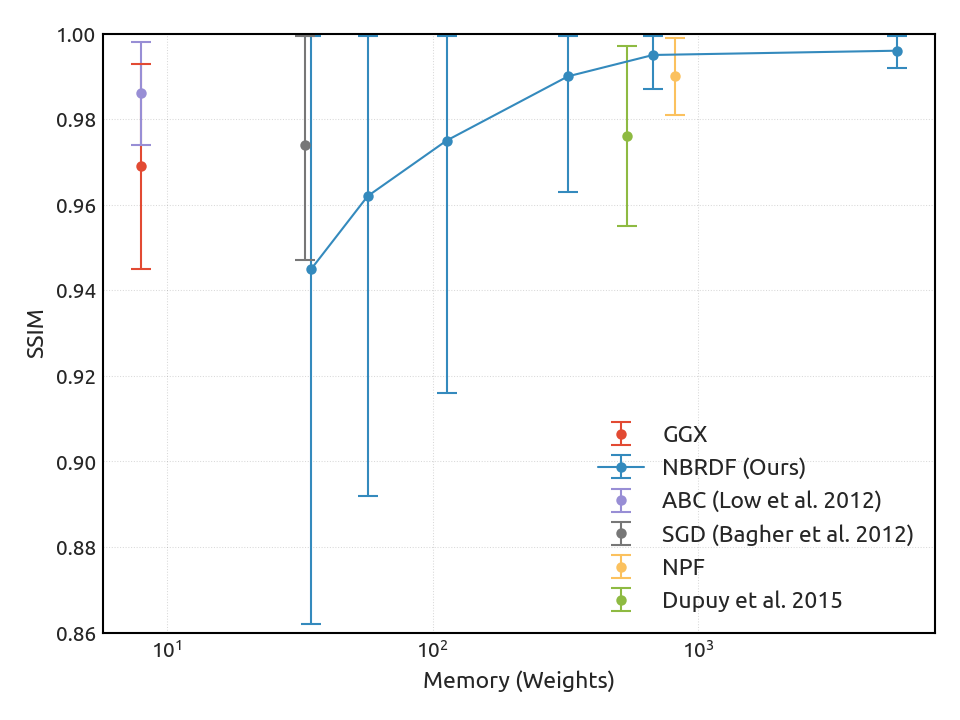}
 \caption{\label{fig:performance}%
 Average SSIM vs Memory footprint (log scale) for multiple representations of BRDFs, with standard deviations. For NBRDFs the network size can be adjusted to select the reconstruction accuracy, thus we include data points for multiple sizes.}
\end{figure}

Finally, in Figure~\ref{fig:performance} we compare multiple BRDF representation methods in terms of the average reconstruction
SSIM error in the MERL database, and the memory footprint of the encoding. We show that the NBRDF network size can be adjusted
to select the reconstruction accuracy. For very small networks ($~100$ weights) the NBRDF reconstruction is inaccurate, and
thus parametric representations are to be preferred. However, for NBRDF networks of $~300$ weights the reconstruction
accuracy is already better than the best parametric encoding (Low~\etal~\cite{low2012}) and equivalent to a state-of-the-art
non-parametric method (NPF~\cite{bagher2016}).

\section{Conclusions}
\label{sec:conc}
We propose a compact, accurate neural representation to encode real-world isotropic and anisotropic measured BRDFs.
Combining the learning power of neural networks with a continuous parametrisation allows us to train a representation that implicitly interpolates, and preserves fidelity to the original data at high compression rates.
A new network instance is trained for every new material, but the training is fast and efficient as the networks are very light-weight.

We also show that the models are sufficiently well behaved to be further compressed by an autoencoder.
The learned embedding space of materials open doors to new applications such as interpolating between materials, and learning to predict material-related properties.
Specifically, we show that the latent positions can be mapped to importance sampling parameters of a given distribution.
The computational cost of network evaluation is not significantly higher than equivalent analytic BRDFs, and the added importance sampling routine allows us to get comparable rendering convergence speed.
Overall, our model provides a high-accuracy real-world BRDF representation, at a rendering performance comparable to analytic models.

In future work, our architecture could be applied to spatially-varying materials,
for instance to derive spatially-varying importance sampling parameters on-the-fly, for procedurally created objects and materials.
Similarly to the importance sampling parameter prediction, our meta-learning architecture
can be used to learn further mappings, enabling applications such as perceptual material editing,
and fast analytic model fitting.

\section{Acknowledgements}
\label{sec:ack}
This project has received funding from the European Union's Horizon 2020 research and innovation program under the Marie Sk{\l}odowska-Curie grant agreement No. 642841 (DISTRO ITN).


\printbibliography                

@inproceedings{sztrajman2017,
  author = {Sztrajman, Alejandro and K\v{r}iv\'anek, Jaroslav and Wilkie, Alexander and Weyrich, Tim},
  title = {Image-based Remapping of Material Appearance},
  booktitle = {Proc. 5th Workshop on Material Appearance Modeling},
  editor = {Reinhard Klein and Holly Rushmeier},
  series = {MAM '17},
  year = 2017,
  month = jun,
  day = 18,
  location = {Helsinki, Finland},
  pages = {5--8},
  publisher = {The Eurographics Association},
  address = {Aire-la-Ville, Switzerland, Switzerland},
  issn = {2309-5059},
  isbn = {978-3-03868-035-2},
  doi = {10.2312/mam.20171323}
}

@article{sztrajman2019,
    author = {Sztrajman, Alejandro and K\v{r}iv\'anek, Jaroslav and Wilkie, Alexander and Weyrich, Tim},
    title = {Image-based Remapping of Spatially-Varying Material Appearance},
    year = 2019,
    month = oct,
    day = 31,
    journal = {Journal of Computer Graphics Techniques (JCGT)},
    volume = 8,
    number = 4,
    pages = {1--30},
    url = {http://jcgt.org/published/0008/04/01/},
    issn = {2331-7418}
}

@inproceedings{ngan2005,
 author = {Ngan, Addy and Durand, Fr{\'e}do and Matusik, Wojciech},
 title = {Experimental Analysis of {BRDF} Models},
 booktitle = {Proceedings of the Sixteenth Eurographics Conference on Rendering Techniques},
 series = {EGSR '05},
 year = {2005},
 isbn = {3-905673-23-1},
 location = {Konstanz, Germany},
 pages = {117--126},
 numpages = {10},
 url = {http://dx.doi.org/10.2312/EGWR/EGSR05/117-126},
 doi = {10.2312/EGWR/EGSR05/117-126},
 acmid = {2383671},
 publisher = {Eurographics Association},
 address = {Aire-la-Ville, Switzerland, Switzerland},
}

@article{ward1992,
 author = {Ward, Gregory J.},
 title = {Measuring and Modeling Anisotropic Reflection},
 journal = {SIGGRAPH Comput. Graph.},
 issue_date = {July 1992},
 volume = {26},
 number = {2},
 month = jul,
 year = {1992},
 issn = {0097-8930},
 pages = {265--272},
 numpages = {8},
 url = {http://doi.acm.org/10.1145/142920.134078},
 doi = {10.1145/142920.134078},
 acmid = {134078},
 publisher = {ACM},
 address = {New York, NY, USA},
}

@article{cooktorrance1982,
 author = {Cook, R. L. and Torrance, K. E.},
 title = {A Reflectance Model for Computer Graphics},
 journal = {ACM Trans. Graph.},
 issue_date = {Jan. 1982},
 volume = {1},
 number = {1},
 month = jan,
 year = {1982},
 issn = {0730-0301},
 pages = {7--24},
 numpages = {18},
 url = {http://doi.acm.org/10.1145/357290.357293},
 doi = {10.1145/357290.357293},
 acmid = {357293},
 publisher = {ACM},
 address = {New York, NY, USA},
}

@article{bieron2020,
author = {Bieron, J. and Peers, P.},
title = {An Adaptive BRDF Fitting Metric},
journal = {Computer Graphics Forum},
volume = {39},
number = {4},
pages = {59-74},
doi = {10.1111/cgf.14054},
url = {https://onlinelibrary.wiley.com/doi/abs/10.1111/cgf.14054},
eprint = {https://onlinelibrary.wiley.com/doi/pdf/10.1111/cgf.14054},
year = {2020}
}

@article{ashikhmin2000anisotropic,
 author = {Ashikhmin, Michael and Shirley, Peter},
 title = {An Anisotropic Phong BRDF Model},
 journal = {J. Graph. Tools},
 issue_date = {Dec. 2000},
 volume = {5},
 number = {2},
 month = feb,
 year = {2000},
 issn = {1086-7651},
 pages = {25--32},
 numpages = {8},
 url = {http://dx.doi.org/10.1080/10867651.2000.10487522},
 doi = {10.1080/10867651.2000.10487522},
 acmid = {358644},
 publisher = {A. K. Peters, Ltd.},
 address = {Natick, MA, USA},
}

@inproceedings{walter2007microfacet,
 author = {Walter, Bruce and Marschner, Stephen R. and Li, Hongsong and Torrance, Kenneth E.},
 title = {Microfacet Models for Refraction Through Rough Surfaces},
 booktitle = {Proceedings of the 18th Eurographics Conference on Rendering Techniques},
 series = {EGSR'07},
 year = {2007},
 isbn = {978-3-905673-52-4},
 location = {Grenoble, France},
 pages = {195--206},
 numpages = {12},
 url = {http://dx.doi.org/10.2312/EGWR/EGSR07/195-206},
 doi = {10.2312/EGWR/EGSR07/195-206},
 acmid = {2383874},
 publisher = {Eurographics Association},
 address = {Aire-la-Ville, Switzerland, Switzerland},
}

@inproceedings{lafortune1997,
 author = {Lafortune, Eric P. F. and Foo, Sing-Choong and Torrance, Kenneth E. and Greenberg, Donald P.},
 title = {Non-linear Approximation of Reflectance Functions},
 booktitle = {Proceedings of the 24th Annual Conference on Computer Graphics and Interactive Techniques},
 series = {SIGGRAPH '97},
 year = {1997},
 isbn = {0-89791-896-7},
 pages = {117--126},
 numpages = {10},
 url = {http://dx.doi.org/10.1145/258734.258801},
 doi = {10.1145/258734.258801},
 acmid = {258801},
 publisher = {ACM Press/Addison-Wesley Publishing Co.},
 address = {New York, NY, USA},
}

@misc{mitsuba,
   Author = {Wenzel Jakob},
   Year = {2010},
   Note = {http://www.mitsuba-renderer.org},
   Title = {Mitsuba renderer}
}

@inproceedings{rusinkiewicz1998,
  author = "Szymon Rusinkiewicz",
  title = "A New Change of Variables for Efficient {BRDF} Representation",
  booktitle = "Rendering Techniques (Proc. Eurographics Workshop on Rendering)",
  year = "1998",
  month = jun
}

@inproceedings{ngan2006,
author = {Ngan, Addy and Durand, Fr{\'e}do and Matusik, Wojciech},
title = {Image-driven Navigation of Analytical BRDF Models},
booktitle = {Proceedings of the 17th Eurographics Conference on Rendering Techniques},
series = {EGSR '06},
year = {2006},
isbn = {3-905673-35-5},
location = {Nicosia, Cyprus},
pages = {399--407},
numpages = {9},
url = {http://dx.doi.org/10.2312/EGWR/EGSR06/399-407},
doi = {10.2312/EGWR/EGSR06/399-407},
acmid = {2383945},
publisher = {Eurographics Association},
address = {Aire-la-Ville, Switzerland, Switzerland},
}

@article{deepbrdf,
author = {Hu, Bingyang and Guo, Jie and Chen, Yanjun and Li, Mengtian and Guo, Yanwen},
year = {2020},
month = {05},
pages = {157-166},
title = {DeepBRDF: A Deep Representation for Manipulating Measured BRDF},
volume = {39},
journal = {Computer Graphics Forum},
doi = {10.1111/cgf.13920}
}

@article{gilles2020,
  title={Unified Neural Encoding of BTFs},
  author={Gilles Rainer and A. Ghosh and Wenzel Jakob and T. Weyrich},
  journal={Comput. Graph. Forum},
  year={2020},
  volume={39},
  pages={167-178}
}

@article{gilles2019,
  author = {Rainer, Gilles and Jakob, Wenzel and Ghosh, Abhijeet and Weyrich, Tim},
  title = {Neural {BTF} Compression and Interpolation},
  journal = {Computer Graphics Forum (Proc. Eurographics)},
  volume = 38,
  number = 2,
  pages = {1--10},
  year = 2019,
  month = may,
  publisher = {Eurographics Association}
}

@article{dong2019,
author = {Dong, Yue},
year = {2019},
month = {07},
pages = {},
title = {Deep appearance modeling: A survey},
volume = {3},
journal = {Visual Informatics},
doi = {10.1016/j.visinf.2019.07.003}
}

@InProceedings{dam,
author = {Maximov, Maxim and Leal-Taixe, Laura and Fritz, Mario and Ritschel, Tobias},
title = {Deep Appearance Maps},
booktitle = {Proceedings of the IEEE/CVF International Conference on Computer Vision (ICCV)},
month = oct,
year = {2019}
}

@article{soler2018,
author = {Soler, Cyril and Subr, Kartic and Nowrouzezahrai, Derek},
year = {2018},
month = {05},
pages = {135-144},
title = {A Versatile Parameterization for Measured Material Manifolds},
volume = {37},
journal = {Computer Graphics Forum},
doi = {10.1111/cgf.13348}
}

@article{zsolnai2018,
  author    = {Zsolnai-Feh{\'{e}}r, K{\'{a}}roly and
              Wonka, Peter and
              Wimmer, Michael},
  title     = {Gaussian material synthesis},
  journal   = {{ACM} Trans. Graph.},
  volume    = {37},
  number    = {4},
  pages     = {76:1--76:14},
  year      = {2018},
  url       = {https://doi.org/10.1145/3197517.3201307},
  doi       = {10.1145/3197517.3201307},
  bibsource = {dblp computer science bibliography, https://dblp.org}
}

@article{jarosz2009,
author = {Jarosz, Wojciech and Carr, Nathan and Jensen, Henrik},
year = {2009},
month = {04},
pages = {577-586},
title = {Importance Sampling Spherical Harmonics},
volume = {28},
journal = {Comput. Graph. Forum},
doi = {10.1111/j.1467-8659.2009.01398.x}
}

@article{clarberg2005,
author = {Clarberg, Petrik and Jarosz, Wojciech and Akenine-M\"{o}ller, Tomas and Jensen, Henrik Wann},
title = {Wavelet Importance Sampling: Efficiently Evaluating Products of Complex Functions},
year = {2005},
issue_date = {July 2005},
publisher = {Association for Computing Machinery},
address = {New York, NY, USA},
volume = {24},
number = {3},
issn = {0730-0301},
url = {https://doi.org/10.1145/1073204.1073328},
doi = {10.1145/1073204.1073328},
journal = {ACM Trans. Graph.},
month = jul,
pages = {1166–1175},
numpages = {10}
}

@inproceedings{siggraph2010course,
 title = {Importance Sampling for Production Rendering},
 author = {Mark Colbert and Simon Premoze and Guillaume Francois},
 year = {2010},
 booktitle = {ACM SIGGRAPH 2010 Courses},
 series = {SIGGRAPH `10}
}

@article{blinn77,
author = {Blinn, James F.},
title = {Models of Light Reflection for Computer Synthesized Pictures},
year = {1977},
issue_date = {Summer 1977},
publisher = {Association for Computing Machinery},
address = {New York, NY, USA},
volume = {11},
number = {2},
issn = {0097-8930},
url = {https://doi.org/10.1145/965141.563893},
doi = {10.1145/965141.563893},
journal = {SIGGRAPH Comput. Graph.},
month = jul,
pages = {192–198},
numpages = {7}
}

@inproceedings{lawrence2004,
author = {Lawrence, Jason and Rusinkiewicz, Szymon and Ramamoorthi, Ravi},
title = {Efficient BRDF Importance Sampling Using a Factored Representation},
year = {2004},
isbn = {9781450378239},
publisher = {Association for Computing Machinery},
address = {New York, NY, USA},
url = {https://doi.org/10.1145/1186562.1015751},
doi = {10.1145/1186562.1015751},
booktitle = {ACM SIGGRAPH 2004 Papers},
pages = {496–505},
numpages = {10},
location = {Los Angeles, California},
series = {SIGGRAPH '04}
}

@article{bako2017,
author = {Bako, Steve and Vogels, Thijs and Mcwilliams, Brian and Meyer, Mark and Nov\'{a}K, Jan and Harvill, Alex and Sen, Pradeep and Derose, Tony and Rousselle, Fabrice},
title = {Kernel-Predicting Convolutional Networks for Denoising Monte Carlo Renderings},
year = {2017},
issue_date = {July 2017},
publisher = {Association for Computing Machinery},
address = {New York, NY, USA},
volume = {36},
number = {4},
issn = {0730-0301},
url = {https://doi.org/10.1145/3072959.3073708},
doi = {10.1145/3072959.3073708},
journal = {ACM Trans. Graph.},
month = jul,
articleno = {97},
numpages = {14}
}

@article{bako2019,
author = {Bako, Steve and Meyer, Mark and Derose, T. and Sen, Pradeep},
year = {2019},
month = {10},
pages = {527-542},
title = {Offline Deep Importance Sampling for Monte Carlo Path Tracing},
volume = {38},
journal = {Computer Graphics Forum},
doi = {10.1111/cgf.13858}
}

@article{chakravarty2017,
author = {Chaitanya, Chakravarty R. Alla and Kaplanyan, Anton S. and Schied, Christoph and Salvi, Marco and Lefohn, Aaron and Nowrouzezahrai, Derek and Aila, Timo},
title = {Interactive Reconstruction of Monte Carlo Image Sequences Using a Recurrent Denoising Autoencoder},
year = {2017},
issue_date = {July 2017},
publisher = {Association for Computing Machinery},
address = {New York, NY, USA},
volume = {36},
number = {4},
issn = {0730-0301},
url = {https://doi.org/10.1145/3072959.3073601},
doi = {10.1145/3072959.3073601},
journal = {ACM Trans. Graph.},
month = jul,
articleno = {98},
numpages = {12}
}

@article{kallweit2017,
author = {Kallweit, Simon and M\"{u}ller, Thomas and Mcwilliams, Brian and Gross, Markus and Nov\'{a}k, Jan},
title = {Deep Scattering: Rendering Atmospheric Clouds with Radiance-Predicting Neural Networks},
year = {2017},
issue_date = {November 2017},
publisher = {Association for Computing Machinery},
address = {New York, NY, USA},
volume = {36},
number = {6},
issn = {0730-0301},
url = {https://doi.org/10.1145/3130800.3130880},
doi = {10.1145/3130800.3130880},
journal = {ACM Trans. Graph.},
month = nov,
articleno = {231},
numpages = {11}
}

@article{ren2013,
author = {Ren, Peiran and Wang, Jiaping and Gong, Minmin and Lin, Stephen and Tong, Xin and Guo, Baining},
title = {Global Illumination with Radiance Regression Functions},
year = {2013},
issue_date = {July 2013},
publisher = {Association for Computing Machinery},
address = {New York, NY, USA},
volume = {32},
number = {4},
issn = {0730-0301},
url = {https://doi.org/10.1145/2461912.2462009},
doi = {10.1145/2461912.2462009},
journal = {ACM Trans. Graph.},
month = jul,
articleno = {130},
numpages = {12}
}

@article{zheng2019,
author = {Zheng, Quan and Zwicker, Matthias},
year = {2019},
month = {05},
pages = {169-179},
title = {Learning to Importance Sample in Primary Sample Space},
volume = {38},
journal = {Computer Graphics Forum},
doi = {10.1111/cgf.13628}
}

@article{mueller2019,
	author = {Thomas M\"{u}ller and Brian McWilliams and Fabrice Rousselle and Markus Gross and Jan Nov\'{a}k},
	title = {Neural Importance Sampling},
	journal = {ACM Trans. Graph.},
	issue_date = {October 2019},
	volume = {38},
	number = {5},
	month = oct,
	year = {2019},
	issn = {0730-0301},
	pages = {145:1--145:19},
	articleno = {145},
	numpages = {19},
	url = {http://doi.acm.org/10.1145/3341156},
	doi = {10.1145/3341156},
	acmid = {3341156},
	publisher = {ACM},
	address = {New York, NY, USA},
}

@inproceedings{dinh2017,
  author    = {Dinh, Laurent and
               Sohl-Dickstein, Jascha and
               Bengio, Samy},
  title     = {Density estimation using Real {NVP}},
  booktitle = {5th International Conference on Learning Representations, {ICLR} 2017,
               Toulon, France, April 24-26, 2017, Conference Track Proceedings},
  publisher = {OpenReview.net},
  year      = {2017},
  url       = {https://openreview.net/forum?id=HkpbnH9lx},
  bibsource = {dblp computer science bibliography, https://dblp.org}
}

@misc{dinh2015,
      title={NICE: Non-linear Independent Components Estimation}, 
      author={Laurent Dinh and David Krueger and Yoshua Bengio},
      year={2015},
      eprint={1410.8516},
      archivePrefix={arXiv},
      primaryClass={cs.LG}
}

@inproceedings{brdfstar2016,
author = {Guarnera, D. and Guarnera, G. C. and Ghosh, A. and Denk, C. and Glencross, M.},
title = {BRDF Representation and Acquisition},
year = {2016},
publisher = {Eurographics Association},
address = {Goslar, DEU},
booktitle = {Proceedings of the 37th Annual Conference of the European Association for Computer Graphics: State of the Art Reports},
pages = {625–650},
numpages = {26},
location = {Lisbon, Portugal},
series = {EG '16}
}

@article{utiadensebrdf,
title = {{Adaptive slices for acquisition of anisotropic BRDF}},
author = {V{\'a}vra, R. and Filip, J.},
journal = {Computational Visual Media},
issue = {1},
number = {4},
year = {2018},
publisher = {Springer},
DOI = {https://doi.org/10.1007/s41095-017-0099-z},
pages = {55--69}
}

@article{mpbm03,
author = {Matusik, Wojciech and Pfister, Hanspeter and Brand, Matthew and Mcmillan, Leonard},
year = {2003},
month = {07},
pages = {759-769},
title = {A Data-Driven Reflectance Model},
volume = {22},
journal = {ACM Trans. Graph.},
doi = {10.1145/1201775.882343}
}

@article{deschaintre2018,
author = {Deschaintre, Valentin and Aittala, Miika and Durand, Fredo and Drettakis, George and Bousseau, Adrien},
title = {Single-Image SVBRDF Capture with a Rendering-Aware Deep Network},
year = {2018},
issue_date = {August 2018},
publisher = {Association for Computing Machinery},
address = {New York, NY, USA},
volume = {37},
number = {4},
issn = {0730-0301},
url = {https://doi.org/10.1145/3197517.3201378},
doi = {10.1145/3197517.3201378},
journal = {ACM Trans. Graph.},
month = jul,
articleno = {128},
numpages = {15}
}

@article{nielsen2015,
author = {Nielsen, Jannik Boll and Jensen, Henrik Wann and Ramamoorthi, Ravi},
title = {On Optimal, Minimal BRDF Sampling for Reflectance Acquisition},
year = {2015},
issue_date = {November 2015},
publisher = {Association for Computing Machinery},
address = {New York, NY, USA},
volume = {34},
number = {6},
issn = {0730-0301},
url = {https://doi.org/10.1145/2816795.2818085},
doi = {10.1145/2816795.2818085},
journal = {ACM Trans. Graph.},
month = oct,
articleno = {186},
numpages = {11}
}

@article{dupuy2015,
author = {Dupuy, Jonathan and Heitz, Eric and Iehl, Jean-Claude and Poulin, Pierre and Ostromoukhov, Victor},
title = {Extracting Microfacet-Based BRDF Parameters from Arbitrary Materials with Power Iterations},
year = {2015},
issue_date = {July 2015},
publisher = {The Eurographs Association and John Wiley \& Sons, Ltd.},
address = {Chichester, GBR},
volume = {34},
number = {4},
issn = {0167-7055},
journal = {Comput. Graph. Forum},
month = jul,
pages = {21--30},
numpages = {10}
}

@inproceedings{marschner1999imagebased,
author = {Marschner, Stephen R. and Westin, Stephen H. and Lafortune, Eric P. F. and Torrance, Kenneth E. and Greenberg, Donald P.},
title = {Image-Based {BRDF} Measurement Including Human Skin},
year = {1999},
isbn = {321183382X},
publisher = {Eurographics Association},
address = {Goslar, DEU},
booktitle = {Proceedings of the 10th Eurographics Conference on Rendering},
pages = {131--144},
numpages = {14},
location = {Granada, Spain},
series = {EGWR'99}
}

@article{bagher2012,
author = {Bagher, M. M. and Soler, C. and Holzschuch, N.},
title = {Accurate fitting of measured reflectances using a Shifted Gamma micro-facet distribution},
journal = {Computer Graphics Forum},
volume = {31},
number = {4},
pages = {1509--1518},
doi = {10.1111/j.1467-8659.2012.03147.x},
url = {https://onlinelibrary.wiley.com/doi/abs/10.1111/j.1467-8659.2012.03147.x},
eprint = {https://onlinelibrary.wiley.com/doi/pdf/10.1111/j.1467-8659.2012.03147.x},
year = {2012}
}

@article{low2012,
author = {L\"{o}w, Joakim and Kronander, Joel and Ynnerman, Anders and Unger, Jonas},
title = {BRDF Models for Accurate and Efficient Rendering of Glossy Surfaces},
year = {2012},
issue_date = {January 2012},
publisher = {Association for Computing Machinery},
address = {New York, NY, USA},
volume = {31},
number = {1},
issn = {0730-0301},
url = {https://doi.org/10.1145/2077341.2077350},
doi = {10.1145/2077341.2077350},
journal = {ACM Trans. Graph.},
month = feb,
articleno = {9},
numpages = {14}
}

@article{rgl,
author = {Jonathan Dupuy and Wenzel Jakob},
title = {An Adaptive Parameterization for Efficient Material Acquisition and Rendering},
journal = {Transactions on Graphics (Proceedings of SIGGRAPH Asia)},
volume = {37},
number = {6},
pages = {274:1--274:18},
year = {2018},
month = {nov},
doi = {10.1145/3272127.3275059}
}

@article{bagher2016,
author = {Bagher, Mahdi M. and Snyder, John and Nowrouzezahrai, Derek},
title = {A Non-Parametric Factor Microfacet Model for Isotropic BRDFs},
year = {2016},
issue_date = {September 2016},
publisher = {Association for Computing Machinery},
address = {New York, NY, USA},
volume = {35},
number = {5},
issn = {0730-0301},
url = {https://doi.org/10.1145/2907941},
doi = {10.1145/2907941},
journal = {ACM Trans. Graph.},
month = jul,
articleno = {159},
numpages = {16}
}

@misc{keras,
  title={Keras},
  author={Chollet, Fran\c{c}ois and others},
  year={2015},
  howpublished={\url{https://keras.io}},
}

@InProceedings{chen2020,
author="Chen, Zhe
and Nobuhara, Shohei
and Nishino, Ko",
editor="Vedaldi, Andrea
and Bischof, Horst
and Brox, Thomas
and Frahm, Jan-Michael",
title="Invertible Neural BRDF for Object Inverse Rendering",
booktitle="Computer Vision -- ECCV 2020",
year="2020",
publisher="Springer International Publishing",
address="Cham",
pages="767--783",
isbn="978-3-030-58558-7"
}

@InProceedings{romeiro2008,
author="Romeiro, Fabiano
and Vasilyev, Yuriy
and Zickler, Todd",
editor="Forsyth, David
and Torr, Philip
and Zisserman, Andrew",
title="Passive Reflectometry",
booktitle="Computer Vision -- ECCV 2008",
year="2008",
publisher="Springer Berlin Heidelberg",
address="Berlin, Heidelberg",
pages="859--872",
isbn="978-3-540-88693-8"
}

@inproceedings{lindell2021,
author = {Lindell, David B. and Martel, Julien N.P. and Wetzstein, Gordon},
title = {AutoInt: Automatic Integration for Fast Neural Volume Rendering},
booktitle = {Proceedings of the conference on Computer Vision and Pattern Recognition (CVPR)},
year={2021}
}

\end{document}